\documentclass[twocolumn,times]{aastex63}
\usepackage{indentfirst}
\usepackage{graphicx}
\usepackage{mathrsfs}
\usepackage{amsmath}	
\usepackage{amssymb}	
\usepackage[T1]{fontenc}
\usepackage{ae,aecompl}
\usepackage{amsmath}	
\usepackage{amssymb}	
\usepackage{txfonts}

\usepackage{CJKutf8}

\definecolor{color0}  {RGB}{50,150,120}

\newcommand{\kcc}[1]{{\color{black}#1}}
\newcommand{\cntext}[1]{\begin{CJK*}{UTF8}{bsmi}#1\end{CJK*}}

\usepackage{natbib}
\received{\today}
\submitjournal{ApJ}

\shorttitle{Search for Surviving Companions}
\shortauthors{Rau et al.}

\begin{document}

\title{Evolution of Main-Sequence-like Surviving Companions in Type Ia Supernova Remnants}
\author[0000-0003-4692-5941]{Shiau-Jie Rau (\cntext{饒孝節})}
\affiliation{Department of Physics, National Tsing Hua University, Hsinchu 30013, Taiwan}
\affiliation{Institute of Astronomy, National Tsing Hua University, Hsinchu 30013, Taiwan}
\affiliation{Center for Informatics and Computation in Astronomy, National Tsing Hua University, Hsinchu 30013, Taiwan}
\affiliation{Physics Division, National Center for Theoretical Sciences, Taipei 10617, Taiwan}

\author[0000-0002-1473-9880]{Kuo-Chuan Pan (\cntext{潘國全})}
\affiliation{Department of Physics, National Tsing Hua University, Hsinchu 30013, Taiwan}
\affiliation{Institute of Astronomy, National Tsing Hua University, Hsinchu 30013, Taiwan}
\affiliation{Center for Informatics and Computation in Astronomy, National Tsing Hua University, Hsinchu 30013, Taiwan}
\affiliation{Center for Theory and Computation, National Tsing Hua University, Hsinchu 30013, Taiwan}


\begin{abstract}

Recent theoretical and numerical studies of Type Ia supernova explosion within the single-degenerate scenario suggest that the non-degenerate companions could survive \kcc{during} the supernova impact and could be detectable in nearby supernova remnants. However, observational efforts show less promising evidence on the existence of surviving companions from the standard single-degenerate channels. 
\kcc{The spin-up/spin-down models are possible mechanisms to explain the non-detection of surviving companions.}
\kcc{In these models, the spin-up phase could increase the critical mass for explosion, leading to a super-Chandrasekhar mass explosion, 
and the spin-down phase could lead to extra mass loss and angular momentum redistribution. }
Since the spin-down timescale for the delayed explosion of a rotating white dwarf is unclear, in this paper, we explore a vast parameter space of main-sequence-like surviving companions via two-dimensional hydrodynamic simulations of supernova impact and the subsequent stellar evolution of surviving companions. 
\kcc{Tight universal relations to describe the mass stripping effect, supernova kick, and depth of supernova heating are provided.} 
Our results suggest that the not-yet detected surviving companions from observations of nearby Type Ia supernova remnants might favor low mass companions, 
short binary separation, or stronger supernova explosion energies than the standard singe-degenerate channels.

\end{abstract}


%
%
%
\section{Introduction}

Type Ia supernovae (SNe~Ia) are famous for their usage as ``standardizable candles'' for cosmological applications \citep{BT92, Riess98, Schmidt98, Perlmutter99}, 
albeit their progenitor systems and intrinsic variations remain unclear \citep{2013FrPhy...8..116H, 2018PhR...736....1L, Rev}.  
Recent progress on time-domain astronomy reveals several variants of peculiar SNe Ia \citep{2019NatAs...3..706J}, 
including 91T-like \citep{2012MNRAS.425.1917S}, 91bg-like \citep{1992AJ....104.1543F}, 02es-like \citep{2012ApJ...751..142G, 2021ApJ...919..142B}, 
SNe Ia CSM \citep{2013ApJS..207....3S}, SNe Iax \citep{2013ApJ...767...57F}, ...etc., 
suggesting multiple channels could produce SNe Ia. 

Observational evidence suggests that SNe Ia must come from thermonuclear explosions of carbon-oxygen white dwarfs in binary systems \citep{2013FrPhy...8..116H}. 
The binary companion could be either another degenerate white dwarf, the double degenerate scenario (DDS; \citealt{1984ApJS...54..335I, 1984ApJ...277..355W}), 
or a non-degenerate companion, such as main-sequence (MS) stars, helium stars, or red giants \citep{Rev}, the Single-Degenerate Scenario (SDS; \citealt{WI73, N82}).  
One major theoretical prediction to distinguish between the DDS and the SDS is that a surviving companion is expected in the SDS through several multi-dimensional hydrodynamics calculations on the interaction of supernova ejecta with a non-degenerate companion \citep{2010ApJ...715...78P, 2012ApJ...750..151P, 2012ApJ...760...21P, 2013ApJ...773...49P, 2013ApJ...778..121L, 2014ApJ...792...71P, 2019ApJ...887...68B, 2020ApJ...898...12Z, 2021MNRAS.500..301L}.
Direct observation of a surviving companion could link the progenitor channel to the explosion environment, which would place meaningful constraints on the intrinsic properties of SNe Ia.   

Numerous observational attempts to search for surviving companions in Type Ia supernova remnants (Ia SNRs) have been made in the past decade, 
including searches in galactic Ia SNRs, e.g., Tycho's SNR \citep{2004Natur.431.1069R, 2009ApJ...691....1G, 2013ApJ...774...99K}, 
Keper's SNR \citep{2018ApJ...862..124R}, and
SN1006 \citep{2012Natur.489..533G, 2018MNRAS.479..192K}, 
and extra-galactic Ia SNRs in the Large Magellanic Cloud \citep{2012Natur.481..164S, 2015ApJ...799..101P, 2017ApJ...837..111L, 2017ApJ...836...85L, Liobs}. 
A few abnormal stars have been found, but no conclusive evidence for the existence of surviving companions \citep{Rev}. 

Theorists have been puzzling about the non-detection of surviving companions for more than a decade. 
The prediction of surviving companions is based on the assumption that the mass transfer from the non-degenerate companion 
is through Roche-lobe overflow (RLOF) or winds in close binary with a binary separation, $a \sim 3R_*$, in the majority of SDS channels, 
where $R_*$ is the companion radius \citep{2012ApJ...750..151P, 2013ApJ...778..121L, 2019ApJ...887...68B}. 

In order to avoid high mass transfer rates during the binary evolution, which might end up in a common envelope instead of a supernova, 
the optically thick wind model \citep{1996ApJ...470L..97H} and 
the common envelope wind model \citep{2017MNRAS.469.4763M} are successful models 
to explain the observed SN Ia rates and allow more massive companions in the initial configuration. 
For instance, in the MS channel in \cite{2008ApJ...679.1390H}, the allowed companion mass at the onset of the SN Ia explosion is about $1.5 - 2M_\odot$.
However, in \cite{2012ApJ...760...21P}, the predicted surviving companions should still be detectable in nearby Ia SNRs ($L \sim 10 - 300 L_\odot$). 

\cite{2011ApJ...738L...1D} proposes that if the mass transfer from binary companion carries angular momentum, 
the accreting white dwarf could be spin up and increase the Chandrasekhar mass \kcc{for explosion}. 
\kcc{If the rotating and accreting WD could reach the new Chandrasekhar mass, a super-Chandrasekhar mass explosion could happen,
otherwise, it takes extra time to spin down. }
The explosion will be delayed after an unknown spin-down period. 
During this spin-down phase, the binary configuration could be significantly changed depending on the spin-down timescale.
\kcc{For instance, extra mass loss of the companion star and re-distribution of orbital angular momentum could happen \citep{2011ApJ...738L...1D,2012ApJ...759...56D}.}
Therefore the prediction of surviving companions could be very different as well \citep{2010MNRAS.401.2729W, 2012ApJ...759...56D, 2021MNRAS.507.4603M}. 
\kcc{It should be noted that this spin-up/down model still have large uncertainty.} 
\kcc{Thus, we use a parameterized method to mimic this effect.}

In this paper, we relax the assumption of RLOF and conduct a parameter study on the evolution of surviving companions in the main-sequence channel. 
The paper is organized as follows. In Section~\ref{sec2}, we describe our simulation codes and numerical setup. 
In Section~\ref{sec3}, we present the results of our simulations of surviving companions. 
Finally, we summarize our results and conclude in Section~\ref{sec4}.

%
%

\section{Numerical Methods \label{sec2}}

Our simulation setup is essentially similar to what has been described in \cite{2010ApJ...715...78P, 2012ApJ...750..151P, 2012ApJ...760...21P}, 
but with several improvements and updates. 
We divide the simulations into three phases. 
The first phase uses the stellar evolution code {\tt MESA} to construct the non-degenerate 
binary companions at the onset of the explosion. 
The second phase involves two-dimensional hydrodynamics simulations of the 
SN ejecta and the companion interactions using the {\tt FLASH} code. 
The third phase reads the outcome from the hydrodynamics 
simulations as input parameters to perform long-term surviving companion evolutions using {\tt MESA}. 
In the following subsections, we describe the detailed setup of each phase.  

\subsection{MS-like Companions}

We use {\tt MESA} r10398 \citep{MESA11, MESA13, MESA15, MESA18, MESA19} to construct four zero-age MS models 
with masses \kcc{$M_i= 0.8, 1, 1.5$}, and $2M_\odot$ and with metalicity z=0.02. 
Table~\ref{Tab:init_R} summarizes the mass and radius information of our four considered companions. 
Note that the realistic companions should be evolved during the mass transfer stage, as considered in \cite{2012ApJ...760...21P}, 
However, the detailed binary evolution will make the companion structure more complex, 
and it is unclear how the spin-down process will make a difference in the companion structure. 
Therefore, the dependence of companion mass on the surviving companion evolution will be less meaningful than zero-age MS stars. 
For simplicity, we consider zero-age MS companions in this study for a parameter study.  

\begin{deluxetable}{cccc}
    \label{Tab:init_R}
    \tablecaption{Companion models}
    \tablehead{
    \colhead{Abbreviation} & \colhead{M$_{i}$ $[M_\odot]$} & \colhead{R$_*$ $[{\rm cm}]$} & \colhead{$\tau_{dyn}$ $[{\rm sec}]$}  
    }
    \startdata
        0.8M & 0.8 & $4.99 \times 10^{10}$ & 1082 \\
        1M & 1.0 & $6.22 \times 10^{10}$ & 1349 \\
        1.5M & 1.5 & $1.01 \times 10^{11}$ & 2281  \\
        2M & 2.0 & $1.13 \times 10^{11}$ & 2326  \\
    \enddata
    \tablecomments{\kcc{M$_{i}$ is the mass of the companion star right before the SN explosion}, R$_*$ is the companion radius, and \kcc{$\tau_{dyn} = \sqrt{R_*^3/(GM_{i})}$} is the dynamical time scale.}

\end{deluxetable}

%

\subsection{Supernova Impact Simulations}


\begin{deluxetable*}{cccccccccccc}
    \label{Tab:final}
	\tablecaption{
	Simulations. 
	}
	\tablewidth{0pt}
    \tablehead{\colhead{Model} &\colhead{M$_{i}$} &\colhead{ \kcc{$ \frac{a}{R_*}$}} &\colhead{E$_{\rm SN}$} &\colhead{M$_f$} & \colhead{$v_{\rm orb}$} & \colhead{$v_{\rm kick}$} & \colhead{$v_{f}$} & \colhead{\kcc{D$_{\rm heat}$}} & \colhead{\kcc{E$\rm _{heat}$}} &\colhead{M$_{\rm Ni}$}\\
    & $[M_\odot]$ &   & $[E_{W7}]$ &  $[M_\odot]$ & $[$km s$^{-1}]$ & $[$km s$^{-1}]$ & $[$km s$^{-1}]$ &  & [erg] &$[M_\odot]$
    }
  \startdata
		0.8M3R     & 0.8 & 3     & 1   & <0.2788 & >352 & >163 & >388 &  0.559   &1.34$\times$10$^{47}$&  $< 10^{-14}$ \\
		0.8M4R     & 0.8 & 4     & 1   & 0.6233  & 305  & 114  & 325  &  0.734   &4.85$\times$10$^{47}$& $< 10^{-14}$ \\
		0.8M5R     & 0.8 & 5     & 1   & 0.7221  & 273  & 79.4 & 284  &  0.818   &3.95$\times$10$^{47}$& 1.24$\times$10$^{-11}$\\
		0.8M6R     & 0.8 & 6     & 1   & 0.7590  & 249  & 56.6 & 255  &  0.864   &2.89$\times$10$^{47}$& 1.79$\times$10$^{-5}$ \\
		1MRL       & 1   & 2.84  & 1   & 0.6051  & 324  & 116  & 344  &  0.636   &5.78$\times$10$^{47}$& $< 10^{-14}$ \\
		1M3R       & 1   & 3     & 1   & 0.6644  & 315  & 118  & 337  &  0.652   &6.31$\times$10$^{47}$& $< 10^{-14}$ \\
		1M4R       & 1   & 4     & 1   & 0.8610  & 273  & 85.0 & 286  &  0.817   &5.65$\times$10$^{47}$& $< 10^{-14}$ \\
		1M5R       & 1   & 5     & 1   & 0.9340  & 244  & 59.7 & 252  &  0.880   &4.01$\times$10$^{47}$& 1.78$\times$10$^{-10}$\\
		1.5MRL     & 1   & 2.59  & 1   & 1.346   & 266  & 71.1 & 276  &  0.863   &9.46$\times$10$^{47}$& $< 10^{-14}$ \\
		1.5M3R     & 1.5 & 3     & 1   & 1.394   & 247  & 56.2 & 254  &  0.887   &7.60$\times$10$^{47}$& 1.91$\times$10$^{-9}$ \\
		1.5M4R     & 1.5 & 4     & 1   & 1.449   & 214  & 35.6 & 217  &  0.927   &4.51$\times$10$^{47}$& 2.04$\times$10$^{-7}$ \\
		1.5M5R     & 1.5 & 5     & 1   & 1.473   & 192  & 26.1 & 193  &  0.956   &2.85$\times$10$^{47}$& 2.85$\times$10$^{-8}$ \\
		2MRL       & 2   & 2.43  & 1   & 1.825   & 260  & 70.2 & 270  &  0.871   &1.31$\times$10$^{48}$& 5.28$\times$10$^{-12}$\\
		2M3R       & 2   & 3     & 1   & 1.897   & 234  & 49.6 & 240  &  0.919   &9.24$\times$10$^{47}$& 1.02$\times$10$^{-10}$\\
		2M4R       & 2   & 4     & 1   & 1.939   & 203  & 26.8 & 205  &  0.948   &5.24$\times$10$^{47}$& 5.27$\times$10$^{-11}$\\
		2M5R       & 2   & 5     & 1   & 1.976   & 182  & 21.1 & 183  &  0.969   &3.00$\times$10$^{47}$& 7.24$\times$10$^{-8}$ \\
		1M3RE0.1   & 1   & 3     & 0.1 & 0.9565  & 315  & 44.2 & 319  &  0.926   &2.59$\times$10$^{47}$& 3.46$\times$10$^{-4}$ \\
		1M3RE0.5   & 1   & 3     & 0.5 & 0.8262  & 315  & 90.8 & 328  &  0.784   &6.46$\times$10$^{47}$& $< 10^{-14}$ \\
		1M3RE2     & 1   & 3     & 2   & <0.2814 & 315  & 141  & 346  &  0.864   &1.10$\times$10$^{47}$& $< 10^{-14}$ \\
		1M3RE5     & 1   & 3     & 5   & <$5.62 \times 10^{-4}$  & 315 & -- & --  & -- & --  & --         \\
		1.5M3RE0.1 & 1.5 & 3     & 0.1 & 1.484   & 247  & 15.5 & 248  &  0.970   &1.43$\times$10$^{47}$& 3.35$\times$10$^{-4}$ \\
		1.5M3RE0.5 & 1.5 & 3     & 0.5 & 1.444   & 247  & 41.6 & 251  &  0.923   &5.01$\times$10$^{47}$& 3.00$\times$10$^{-7}$ \\
		1.5M3RE2   & 1.5 & 3     & 2   & 1.315   & 247  & 77.8 & 259  &  0.836   &1.04$\times$10$^{48}$& $< 10^{-14}$ \\
		1.5M3RE5   & 1.5 & 3     & 5   & 1.110   & 247  & 111  & 271  &  0.717   &1.35$\times$10$^{48}$& $< 10^{-14}$ \\
		1.5M3RE10  & 1.5 & 3     & 10  & <0.7396 & 247  & >139 & >283 &  0.579   &8.19$\times$10$^{47}$& $< 10^{-14}$ \\
	\enddata
    \tablecomments{ 
    M$_f$ is the final mass of the companion star after the SN explosion, 
    $a$ is the binary separation between the companion star and the white dwarf, 
    E$_{SN}$ is the SN explosion energy in units of W7 energies ($E_{W7}=1.233 \times$10$^{51}$erg), 
    $v_{\rm orb}$ is the orbital speed of the companion right before the SN explosion, 
    $v_{\rm kick}$ is the kick velocity due to SN impact, 
    $v_f = \sqrt{v_{\rm orb}^2 + v_{\rm kick}^2 }$ is the final linear velocity of the surviving companion, 
    \kcc{D$\rm _{heat}$ is the heating depth which describe the location of maximum heating in the mass coordinate (see Sec.~\ref{sec:overview}), 
    E$\rm _{heat}$ is the amount of energy deposition in the surviving companion,}     
    and M$_{Ni}$ is the final bound mass of ${^{56}}$Ni onto the companion.}
    
\end{deluxetable*}

Once the companion models are created, we map the density, temperature, and compositions into axisymmetric, two-dimensional grids in a hydrodynamics code, 
{\tt FLASH} (version 4; \citealt{FLASH4, 2008PhST..132a4046D}). 
We consider only hydrogen ($^1$H), helium ($^4$He), and carbon ($^{12}$C) from the companion model in {\tt FLASH}. 
All elements heavier than carbon are approximated as carbon. 
We use the Helmholtz equation of state \citep{helm1, helm2} and the new multiple Poisson solver \citep{2013ApJ...778..181C} for self-gravity with the maximum angular moment for spherical harmonic $l_{\rm max}=80$.  

Since the companion star will receive a kick from the supernova ejecta, 
we consider a simulation box that includes $r < 15 R_*$ and $-20R_* < z < 10R_*$ in axisymmetric  2D cylindrical coordinates with 10 levels of refinement, 
yielding the finest cell spacing of $1/546 R_*$, where $R_*$ is the companion radius (see Table~\ref{Tab:init_R}).
The simulations with \kcc{initial companion mass $M_i = 0.8 M_\odot$}  have a twice bigger simulation box in each dimension due to a higher kick velocity with this companion, 
but we keep the same spatial resolution by increasing one additional level of refinement.   
The ``outflow'' boundary condition is applied in both outer boundaries. 
The companion star is initialized at grid origin and relaxed in the 2D grids for $\sim 10$ dynamical time scales, 
\kcc{$\tau_{dyn}=\sqrt{R_*^3/(GM_i)}$}. 
During the relaxation phase, we artificially damp the velocity by a factor of 0.97 in each timestep. 
After relaxation, a W7-like explosion \citep{2012ApJ...750..151P} is imposed on the positive z-axis at $z = a \times R_*$, where $a$ is the binary separation.  
We use nickel ($^{56}$Ni) as a tracer to represent the supernova ejecta.  
In the W7 model \citep{W7}, the white dwarf mass $M_{W7}$ is 1.378M$_\odot$; total explosion energy $E_{W7}$ is $1.233\times10^{51}$ erg; 
and the averaged ejecta speed $V_{W7}$ is $8.527\times10^8$ cm/s. 
To further consider super- or sub-Chandrasekhar-mass explosion models, we vary the explosion energy from 0.1 to 10 times of the original value in the W7 model. 
The corresponding averaged ejecta speed are scaled as $\sqrt{ E_{SN} / E_{W7} } \times V_{W7}$, where $E_{SN}$ is the explosion energy. 
We ignore the orbital motion and companion's rotation in this study for simplicity. 

%
%


\begin{figure*}
\centering
\plotone{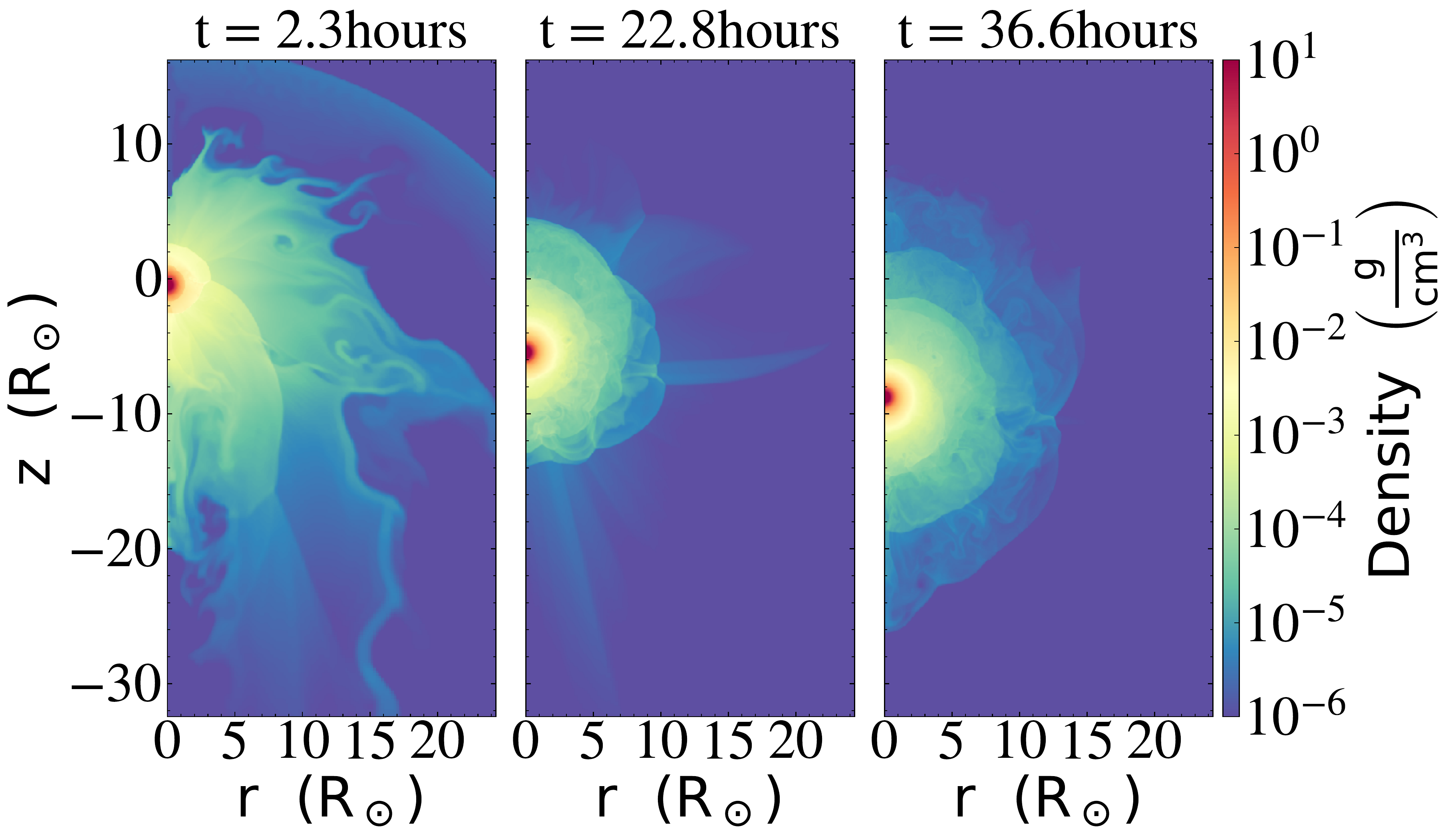}
	\caption{
		Density distributions of a SN impact simulation with model 2M3R. 
		Different panels represent different time after a SN explosion.  
		}
	\label{fig_slice}
\end{figure*}

\subsection{Long-term Evolutions of Surviving Companions \label{sec:sc}}

Since the timestep in {\tt FLASH} simulations is limited by the local sound speed, we could not afford long-term surviving companion evolution within {\tt FLASH}, 
and we do not consider nuclear burning in our simulations. 
Therefore, we terminate the {\tt FLASH} simulations after $\sim 70$ dynamical timescales. Following the method described in \cite{2019ApJ...887...68B}, we measure the stripped mass and entropy changes ($\Delta s$) in our hydrodynamics simulations and assume that the specific entropy will be conserved during the hydrostatic equilibrium process. \kcc{We have tested the entropy changes at different time during the {\tt FLASH} simulation. No significant changes will be appeared in the later on {\tt MESA} simulation.}

We relax the mass of the pre-explode companion models to the final bound mass in the post-impact {\tt FLASH} simulations within {\tt MESA}.
Once the stellar mass is relaxed, we then apply a user-defined heating rate $\dot{\epsilon} = T \Delta s/ \Delta t$ in {\tt MESA} \citep{2019ApJ...887...68B} , 
where $\Delta t$ is the heating timescale and $\Delta t=3$ days. 
After the heating process, we let {\tt MESA} continue running for up to 1 million years to describe the evolution of post-impact surviving companions. 
A convergence test with different heating timescales has been performed. 
We find no significant difference in the long-term evolution after $\gtrsim 10 \Delta t$.

%
%

\section{Results \label{sec3}}

All the simulation parameters we considered in this study are summarized in Table~\ref{Tab:final}. 
For each companion model in Table~\ref{Tab:init_R}, we vary the binary separation from the distance that just allowed Roche-lobe overflow (denoted $R_{\rm RLOF}$) to $5 R_*$ or $6 R_*$ to investigate the separation effects on the evolution of surviving companions. 
In addition, we perform nine extra models with different explosion energies to explore the energy effects.

%
%

\subsection{Overview of the Supernova Impact Simulations \label{sec:overview}}


\begin{figure}
    \centering
	\plottwo{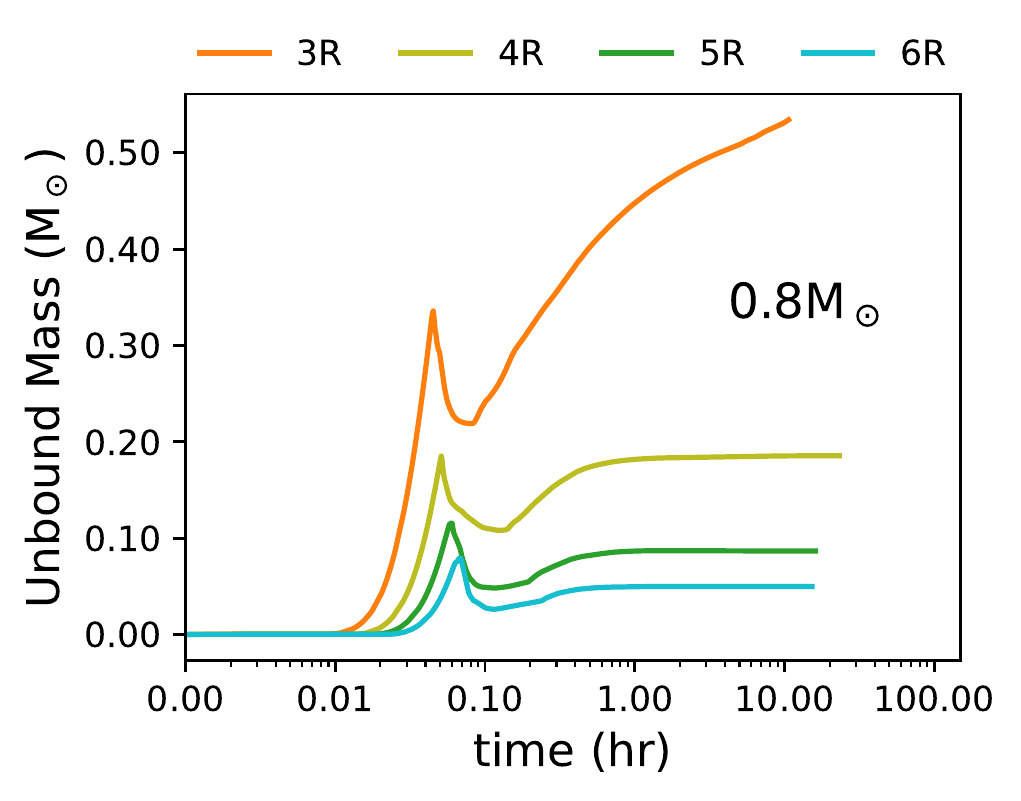}{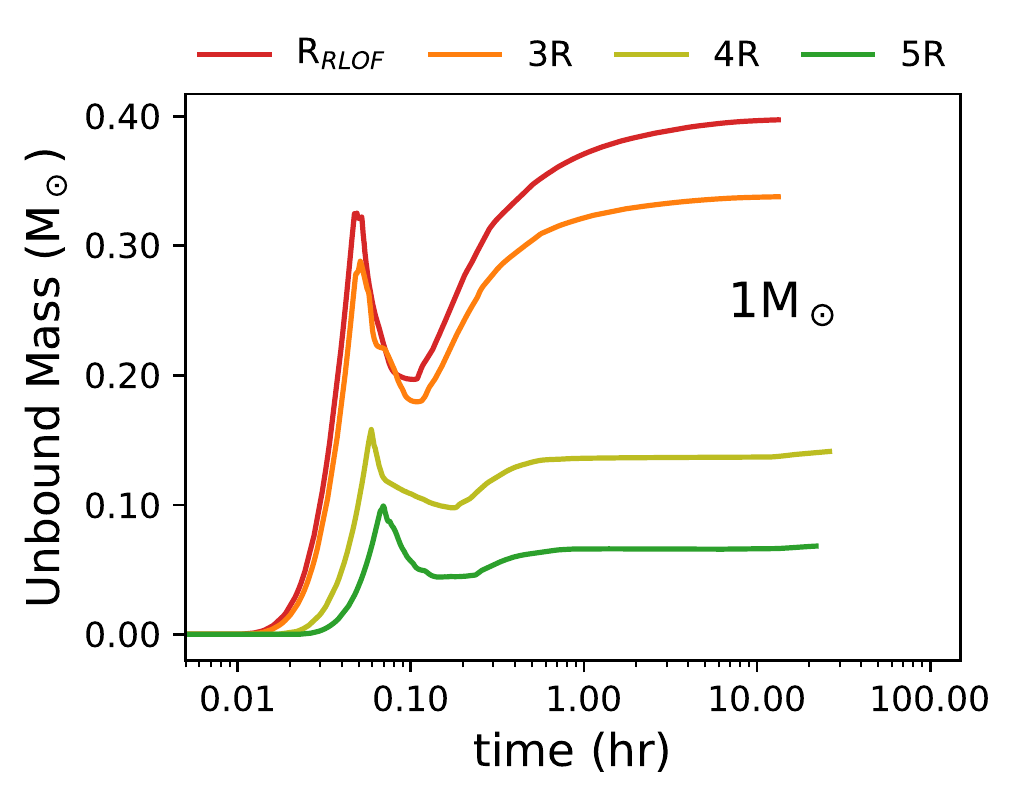}
	\plottwo{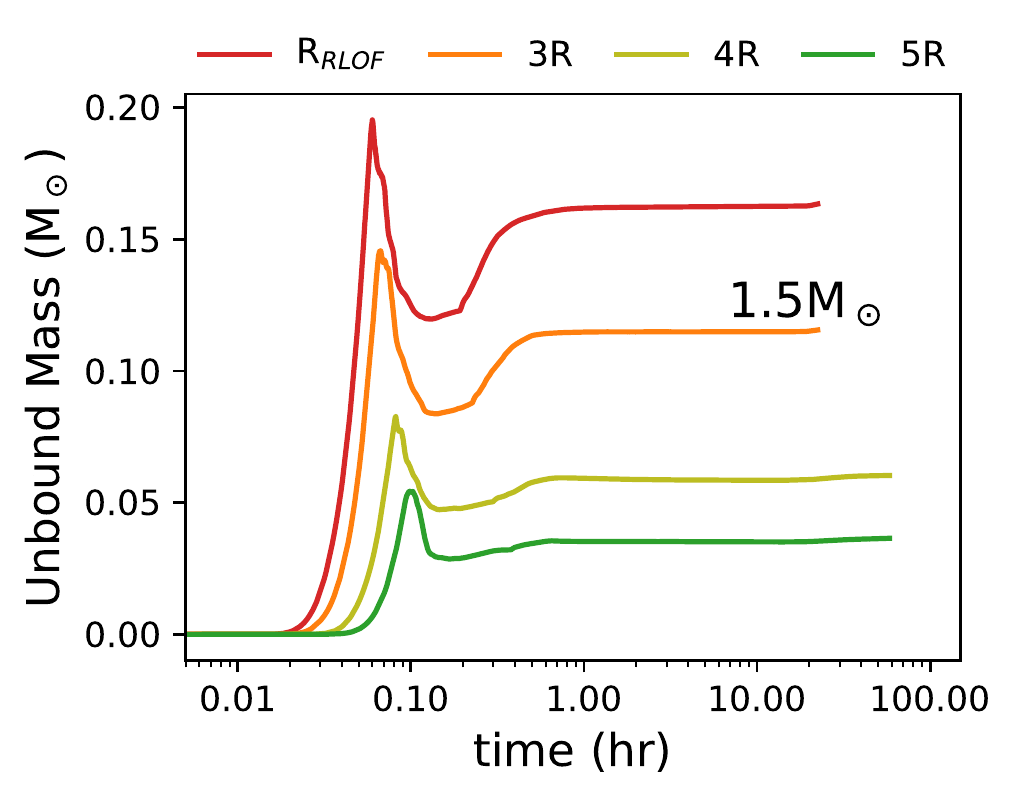}{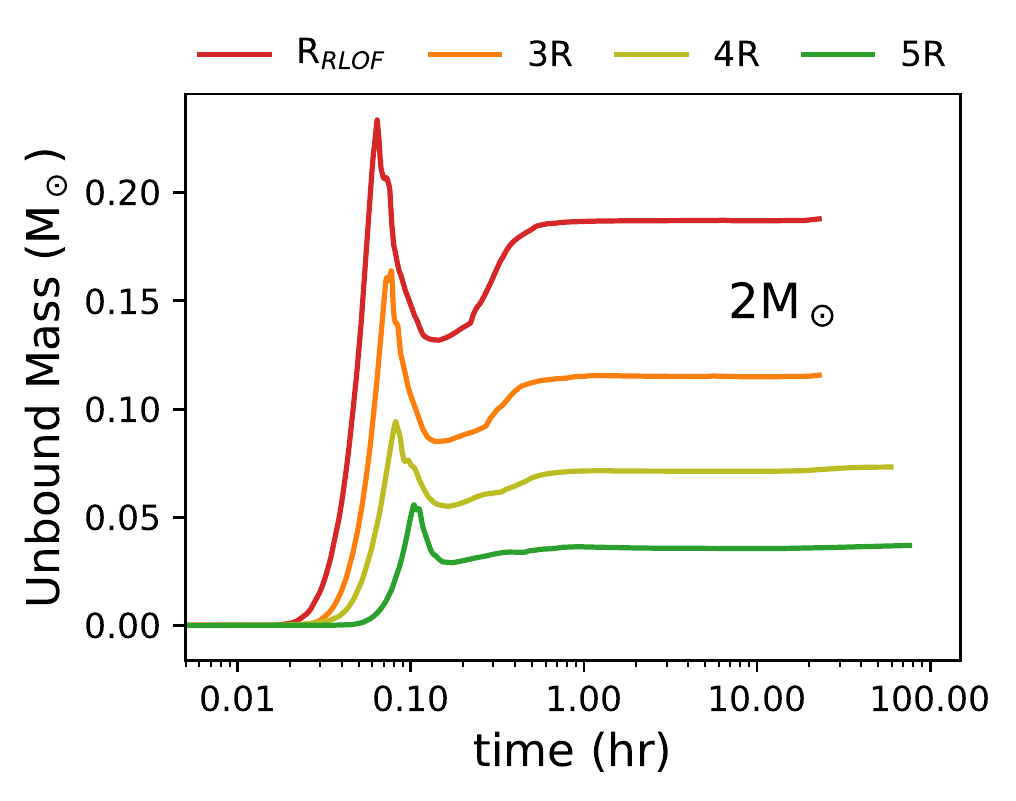}
	\caption{
		Evolutions of the total unbound mass as functions of time. Different panels represent different companion models, and different colors indicate different binary separations (see Table~\ref{Tab:final}).  
	}
	\label{fig_bmass}
\end{figure}

\begin{figure}
    \centering
	\plottwo{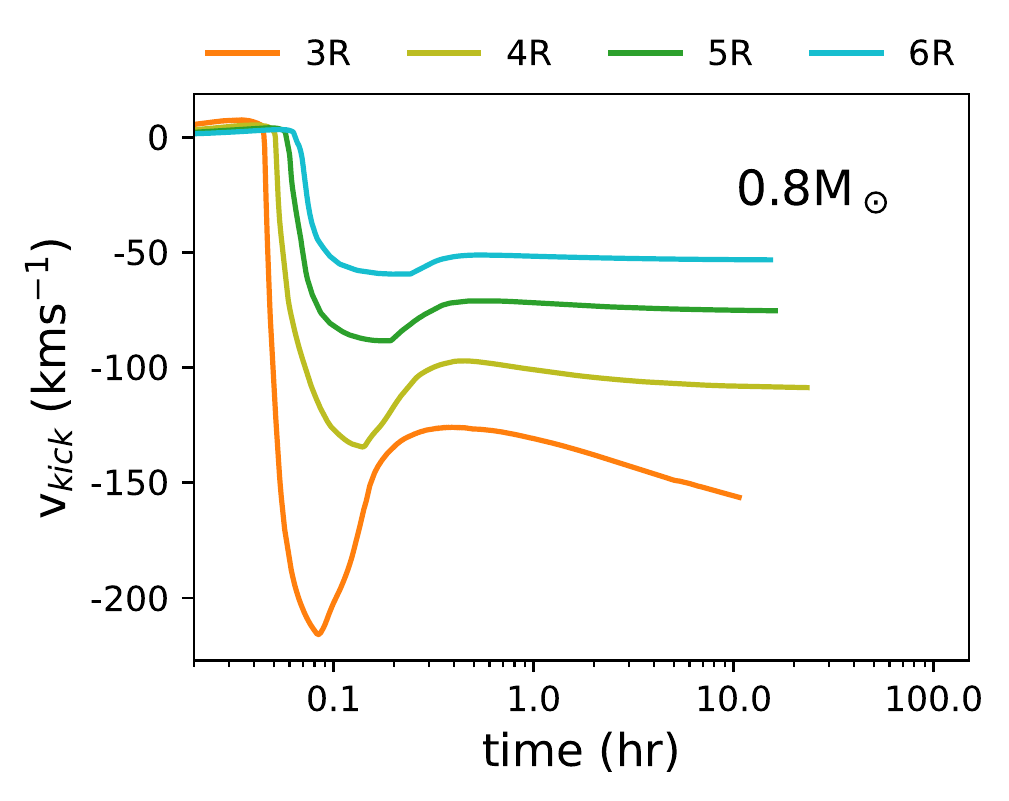}{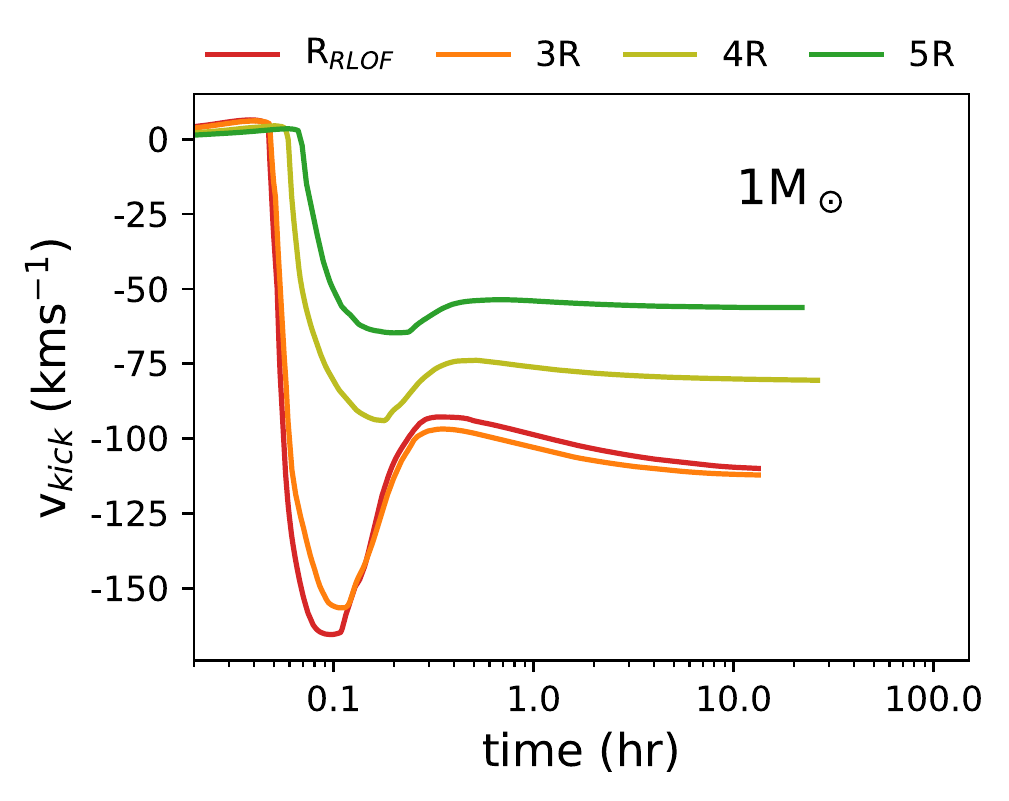}
	\plottwo{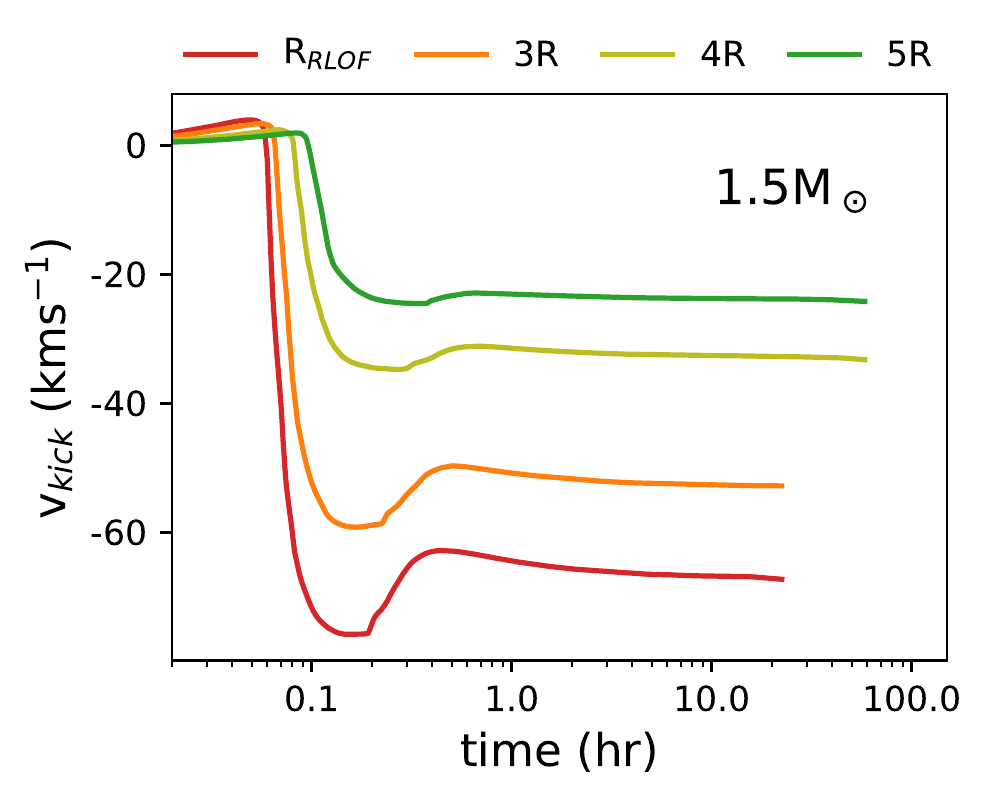}{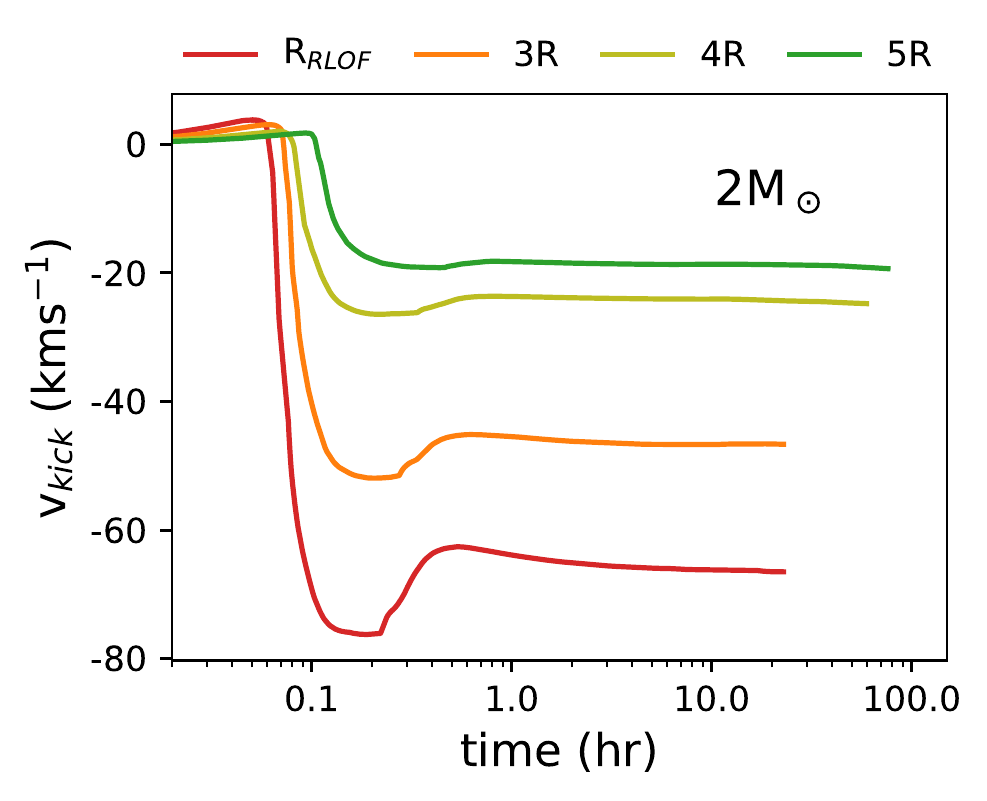}
	\caption{
		Similar to Figure~\ref{fig_bmass} but for the companion's kick velocity. 
	}
	\label{fig_vkick}
\end{figure}

Figure~\ref{fig_slice} shows the density evolution of the model 2M3R (see Table~\ref{Tab:final}) after a SN Ia explosion at different times. 
The overall features are consistent with what has been reported in \cite{2012ApJ...750..151P}. 
Important features that could provide evidence for searches for surviving companions can be summarized as below: 
(1) the collision of supernova eject with the binary companion at around one shock crossing time ($t_{\rm cross} = a / v_{\rm SN} $) 
might emit strong X-ray or UV signals \citep{2010ApJ...708.1025K, 2021ApJ...919..142B}, corresponding to the left panel in Figure~\ref{fig_slice}.  
(2) The ejecta ram pressure (heat) will strip (ablate) a certain amount of mass of the companion star \citep{1975ApJ...200..145W, 2000ApJS..128..615M, 2012ApJ...750..151P}. 
(3) During the collision, the linear momentum of the supernova ejecta could transfer to the surviving companion, resulting in a kick velocity normal to the original orbital velocity \citep{1975ApJ...200..145W, 2000ApJS..128..615M, 2012ApJ...750..151P}. 
(4) The companion star could block the supernova ejecta and form a cone shape cavity behind the companion. 
This cavity might remain open for centuries \citep{2012ApJ...745...75G}.

Figures~\ref{fig_bmass} \& \ref{fig_vkick} show the evolutions of the total unbound companion mass and kick velocity during the supernova impact with different binary separation and companion mass.  
Since the explosions are the same which is described by the W7 model in the previous section, 
lighter companions \kcc{($M_i=0.8$ and $1.0 M_\odot$)} with smaller radii experienced a hotter and stronger impact than massive companions \kcc{($M_i=1.5$ and $2.0 M_\odot$)} with the same \kcc{$a/R_*$} ratio. 
In addition, massive companions have higher gravitational binding energies ($E_b \propto M_i^2/R_*$). 
Therefore, the fractions of final unbound mass with lighter companions are significantly higher than massive companions.

The model 0.8M3R has the closest binary separation to the exploding white dwarf and the lowest gravitational binding energy among all considered companion models. 
We find that this model is almost completely destroyed by the supernova explosion. 
The final bound mass is less than $0.3 M_\odot$ at the end of the simulation, and the kick velocity has reached \kcc{$> 160$~kms$^{-1}$}. 
Although the model 0.8M3R has a twice larger simulation box, the companion reaches the bounding box at $t=10$~hours due to its high kick velocity. 
Therefore, we label the final bound mass (kick velocity) as an upper (lower) limit in Table~\ref{Tab:final}.   

\begin{figure*}
    \centering
        \plottwo{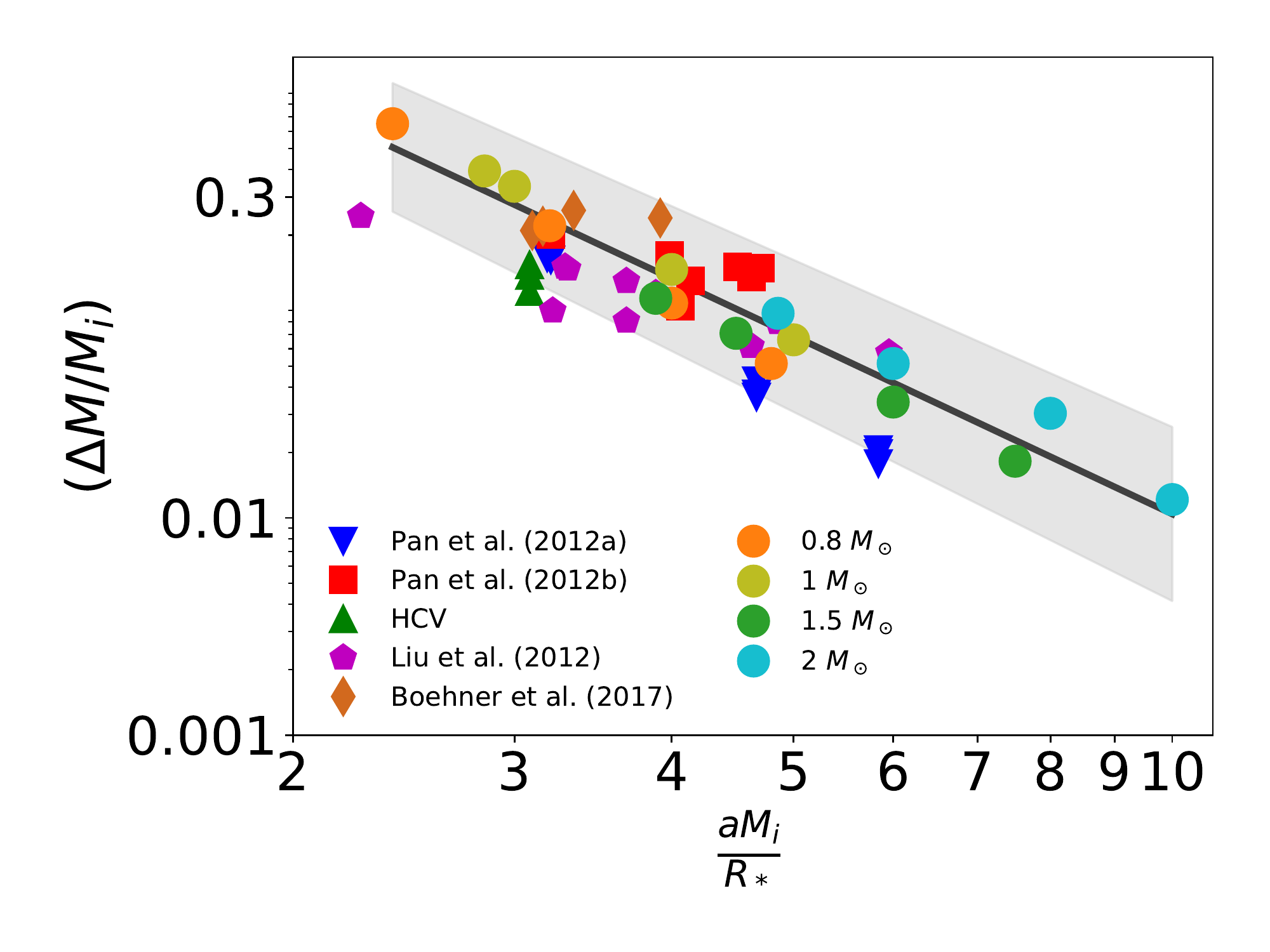}{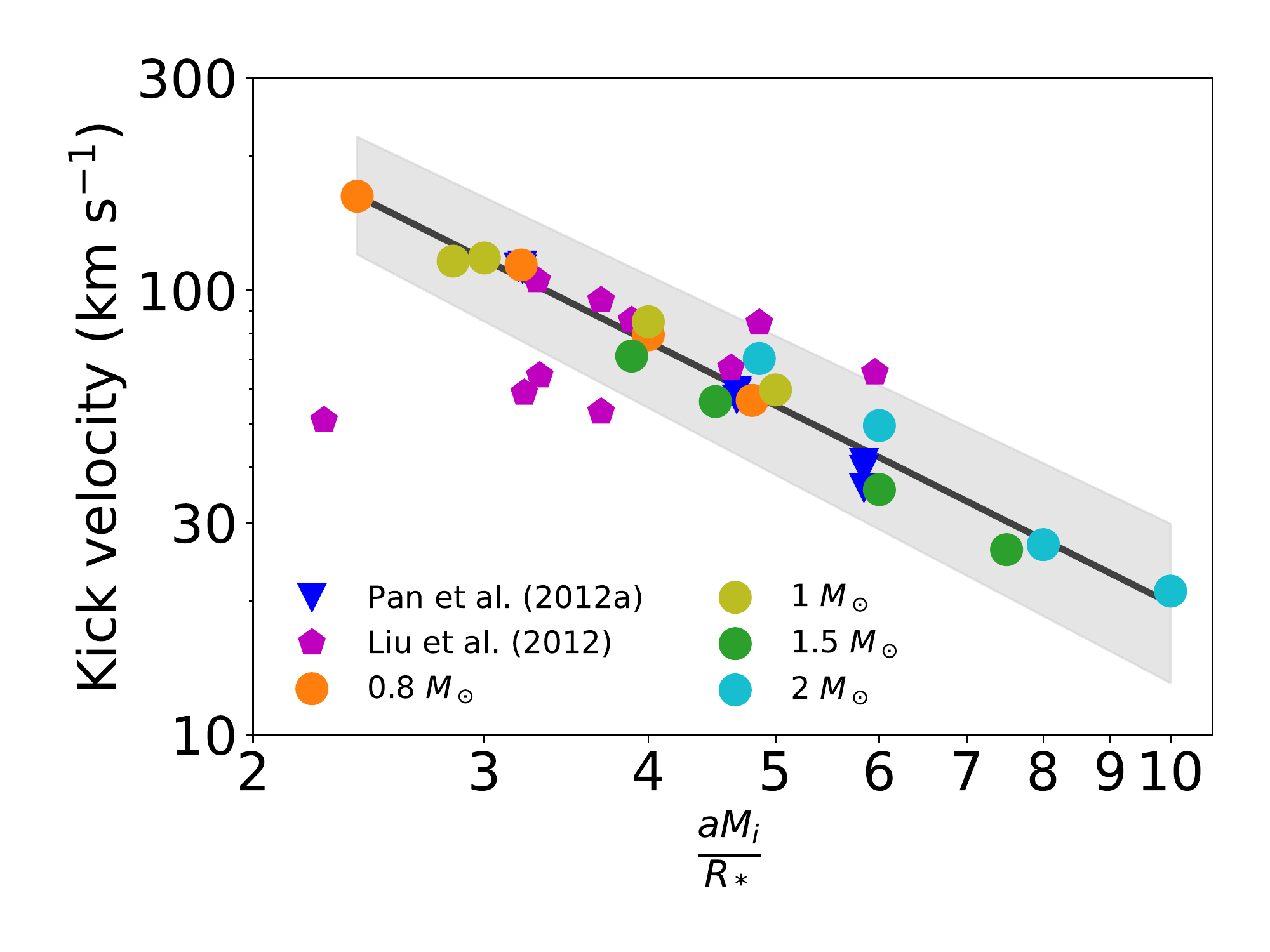}
	\caption{
		Left: the filled circles represent the fraction of the final unbound mass of all models in Table~\ref{Tab:final} with different binary separation. 
		The black line shows a linear fit in log-log scale using {\tt numpy.polyfit}. 
		The gray shaded region indicates a $2\sigma$ confidence interval. 
		Right: similar to the left panel but for the kick velocity. 
	}
	\label{fig_loglogm}
\end{figure*}

\kcc{The relations between the final unbound mass and binary separation can be fitted by a power-law relation, and so does the kick velocity 
(see Figure~\ref{fig_loglogm}). 
The fitted power-law relations are summarized as below:}

\begin{equation}
\frac{\Delta M}{M_{i}} = \left( 5.50^{+1.7}_{-1.3} \right) \times \left(\frac{a M_i}{R_*} \right)^{-2.72 \pm 0.169}, \label{eq_bmass}
\end{equation}

\begin{equation}
v_{\rm kick} = \left( 595.7^{+75.5}_{-67.0} \right) \times \left( \frac{a M_i}{R_*} \right)^{-1.479 \pm 0.0754}~\textrm{[km s}^{-1}\textrm{]}   \label{eq_vkick}.
\end{equation}

Note that, in principle, there are two physical mechanisms to unbound the companion mass, including stripping and ablation \citep{1975ApJ...200..145W, 2012ApJ...750..151P}.
In previous literatures, some authors use the term "stripping mass" to describe the total unbound mass that includes both stripped mass and ablated mass. 
In this paper, we use the term "final unbound mass" to avoid confusion.    

\begin{figure}
	\centering
	\plotone{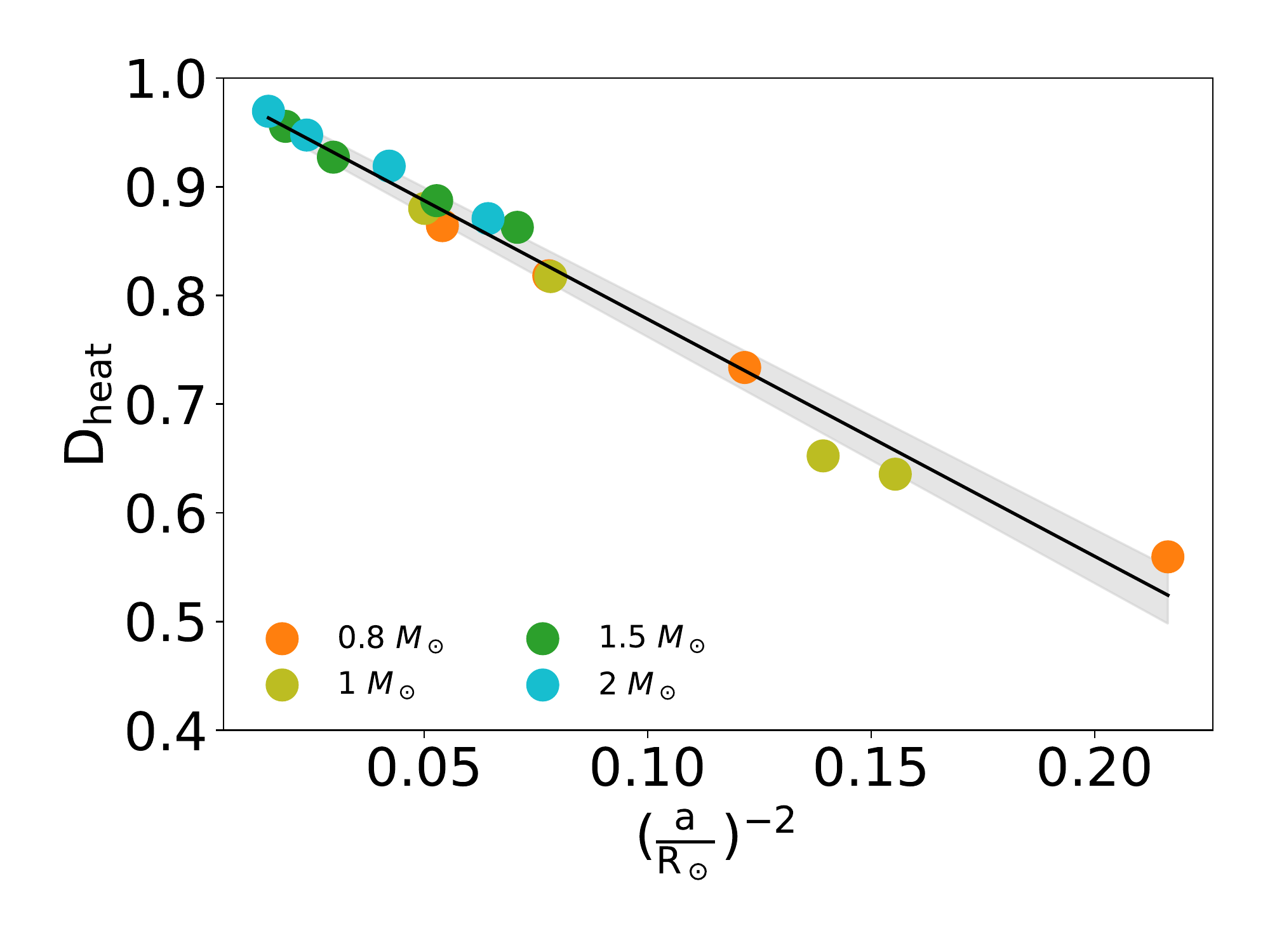}
	\caption{\kcc{Heating depth $D_{\rm heat}$ as functions of one over binary separation square. 
	The x-axis $(\frac{a}{R_\odot})^{-2}$ corresponds to the energy density that hit on the companion star. 
	The black line is the linear fit using {\tt numpy.polyfit}. We label the 2$\sigma$ interval using a gray shaded region.}}
	\label{fig_Dheat}
\end{figure}

The shaded grey band in Figure~\ref{fig_loglogm} represents the one standard deviation of fitted parameters. 
One should note that in previous literatures, the power-law relations are evaluated based on the binary separation ($a$) only. 
The scaling factor, $\frac{M_i}{R_*}$, could be understood by its gravitational binding energy.   
In Figure~\ref{fig_loglogm}, we additionally add data points from the evolved MS companion models in 
\kcc{\cite{2012ApJ...750..151P}, \cite{2012ApJ...760...21P}, \cite{2012A&A...548A...2L}, and \cite{2017MNRAS.465.2060B}}. 
The HCV models (green triangles) are a 1$M_\odot$ MS model that are values carried out by simulations in \cite{2000ApJS..128..615M, SPH:Pakmor08}, and \cite{2012ApJ...750..151P}, using different codes.
The scattering between evolved MS models and ZAMS models in this study could be explained by the changes of stellar structure during the binary evolutions but the overall fitting is quite success, suggesting that this fitted formula could be universal, in first-order of magnitude, for all MS channels. 
\kcc{Note that for 3D hydrodynamics simulations that included the initial orbital motion, only the final linear velocity (including orbital and kick velocity) is recorded. 
Thus, the kick velocities of these simulations are not trivial to extract, and therefore, 
we do not include the data points from \cite{2012ApJ...760...21P} for the kick velocity panel in Figure~\ref{fig_loglogm}.}

\kcc{The post-impact surviving companion receive heating from the SN ejecta as well. 
\cite{2012ApJ...760...21P} suggest that the post-impact evolution of surviving companions depends on both the amount of the SN heating and the depth of energy deposition.
The amount of SN heating is measured by the total heating from the entropy evolution history in {\tt MESA} and summarized in Table~\ref{Tab:final}. 
In general, the models with close binary separation receive a stronger SN energy due to a larger solid angle, but they also suffer a more violent mass striping effect.
The net effect makes no clear relations between the SN heating and binary separation. }

\kcc{However, the location, or the depth, of SN heating is sensitive to the binary separation (see Figure.~\ref{fig_Dheat}). 
Here, we define the heating depth $D_{\rm heat} = m(T_{\rm local~max})/M_f$, 
where $m(T_{\rm local~max})$ is the mass with local temperature maximum due to the SN heating, and $M_f$ is the final companion mass after SN impact.
}

\kcc{A linear relationship between the heating depth and the binary separation can be fitted as bellow,
\begin{equation}
	D_{\rm heat} = -2.183 ^{+0.08}_{-0.08} \times (\frac{a}{R_\odot})^{-2} + 0.996^{+0.08}_{-0.08}. 
	\label{eqn: Dheat}
\end{equation}
\kcc{Note that the heating depth is insensitive to the companion structure of our considered MS companions. }

}

Furthermore, It is possible that part of the supernova ejecta could be bound on the surface of the surviving companion and therefore contaminate its spectrum \citep{2012ApJ...750..151P}. 
This SN contamination could be another smoking gun evidence for searches of surviving companions in SNRs. 
In Table~\ref{Tab:final}, we summarize the final bound nickel mass on the surviving companions.
\kcc{We notice that in some cases with strong SN impact have bound nickel mass closing to the rounding errors ($\sim 10^{-14} M_\odot$). 
Thus, we only list an upper limit of the bound nickel mass in these cases. }
It is found that only low explosion energy cases (models 1M3RE0.1 and 1.5M3RE0.1) or large binary separation cases (0.8M6R) have bound nickel mass $\gtrsim 10^{-5} M_\odot$, 
which are lower than the previous estimation of the nickel contamination from 3D simulations in \cite{2012ApJ...750..151P}.  
\kcc{The lower bound nickel mass in our high-resolution 2D simulations might be due to the stronger turbulent convections around the companion envelope during the SN impact,
comparing to 3D simulations.}

\subsection{Explosion Energy Dependence }

\begin{figure*}
    \centering
	\plottwo{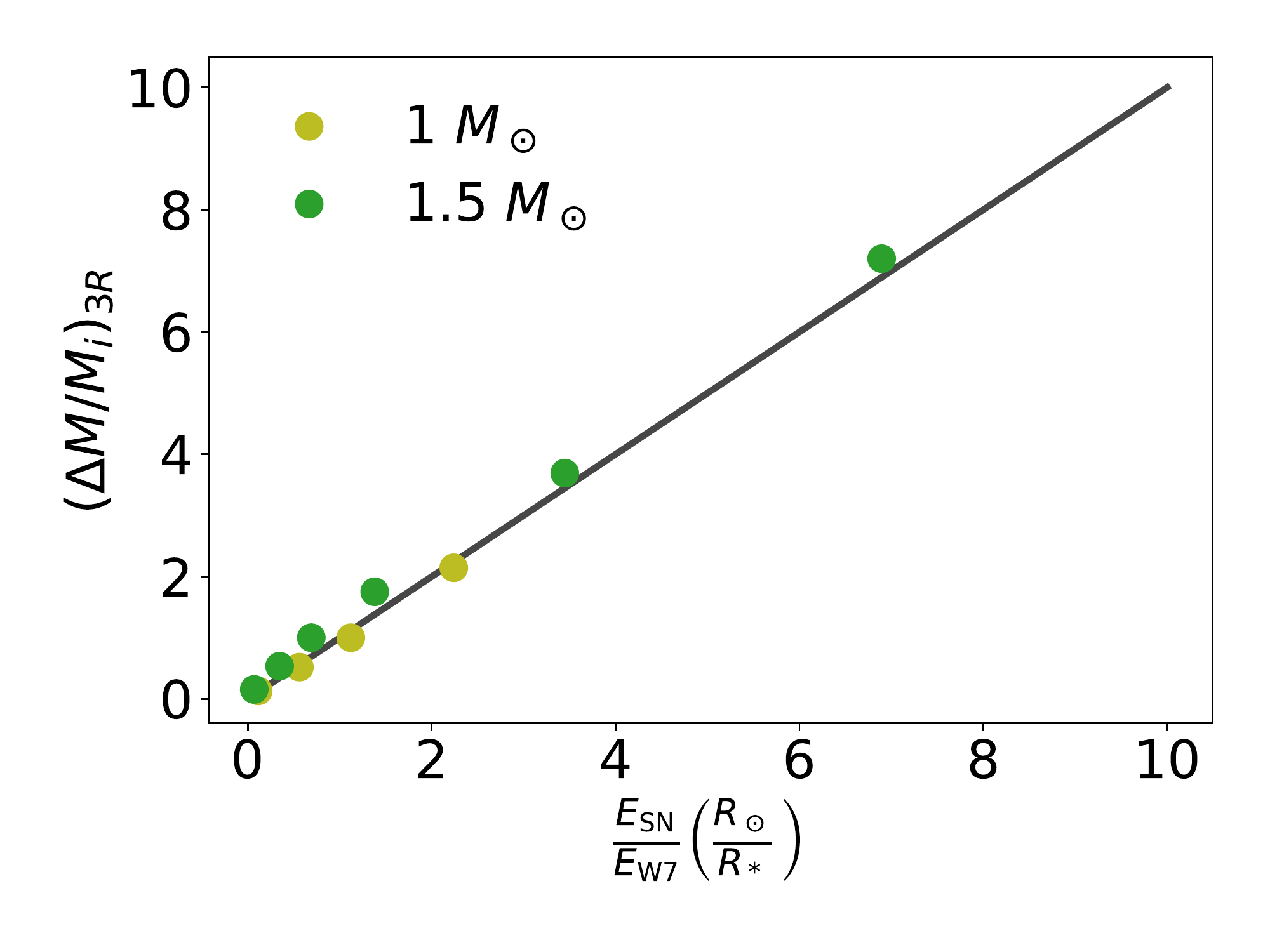}{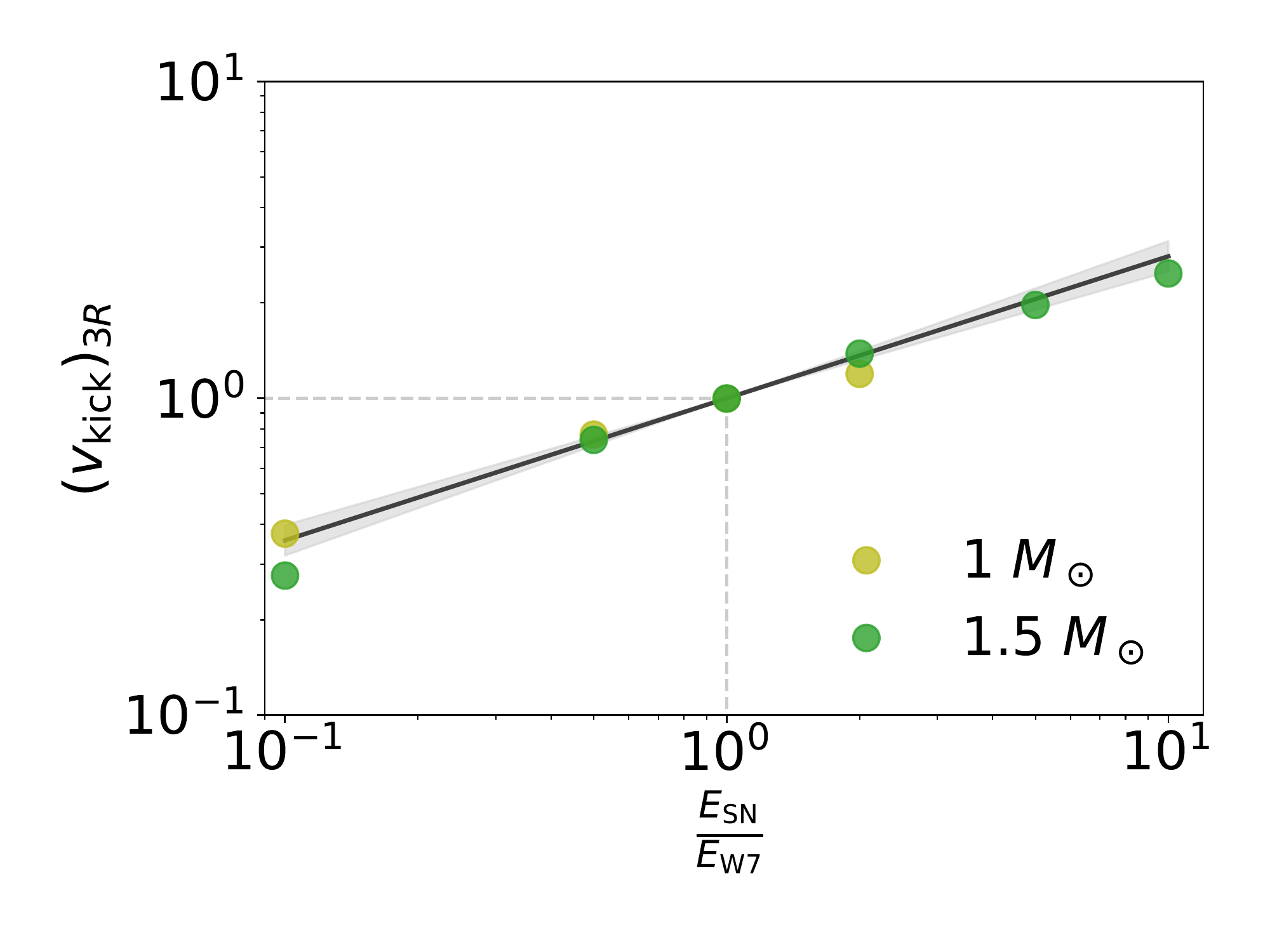}
	\caption{
		Similar to Figure~\ref{fig_loglogm} but for models with different explosion energies. 
	}
	\label{fig_loglog2}
\end{figure*}

The amount of SN explosion energy is another important factor that should affect the conditions of a surviving companion 
but has not been fully discussed in the past (except \citealt{SPH:Pakmor08}). 
The energy difference between sub-luminous SNe~Ia and over-luminous SNe~Ia could be more than an order of magnitude \kcc{\citep{2017hsn..book..317T, 2019NatAs...3..706J}}. 
Figure~\ref{fig_loglog2} shows the influence of SN explosion energies on the final unbound mass and kick velocities with models 1M and 1.5M.  
Similar to what has been reported in \cite{SPH:Pakmor08}, the final unbound mass is linearly proportional to the explosion energies regardless of companion mass, 
and can be described by Equation~\ref{eq:EdM},
\kcc{
\begin{equation}
\left( \frac{\Delta M}{M_i}\right)_{3R_*} = \frac{E_{\rm SN}}{E_{\rm W7}} \left( \frac{R_\odot}{R_*} \right), \label{eq:EdM}
\end{equation}
where $\left( \Delta M / M_i \right)_{3R_*}$ is the percentage of unbound mass normalized to the corresponding model with $a=3R_*$ and $E_{\rm SN} = E_{\rm W7}$, 
and $\Delta M = M_i-M_f$ is the mass difference.
}
In \cite{SPH:Pakmor08}, only one companion model is considered. 
In Equation~\ref{eq:EdM}, we additionally show that the percentage of the unbound mass is \kcc{inversely} proportional to the companion radius.
This could be understood by the fact that the energy received on the companion is $ \propto R_*^2 / a$ and we have chosen $a=3R_*$. 

The right panel in Figure~\ref{fig_loglog2} shows the fraction of kick velocity as functions of explosion energy, 
where \kcc{$\left( v_{\rm kick} \right)_{3R_*}$} is the kick velocity normalized to the corresponding model with $a=3R_*$ and $E_{\rm SN} = E_{\rm W7}$.
The relation could be fitted by a power law in Equation~\ref{eq:EV}.
\kcc{
\begin{equation}
\left( v_{\rm kick} \right)_{3R_*} =  \left(\frac{E_{\rm SN}}{E_{\rm W7}} \right)^{0.4483}, \label{eq:EV}
\end{equation}
}
The power law index 0.448 is less than 0.5, suggesting that a small fraction of ejecta energy is converted to thermal energy by SN heating. 

\begin{figure}
    \centering
	\plottwo{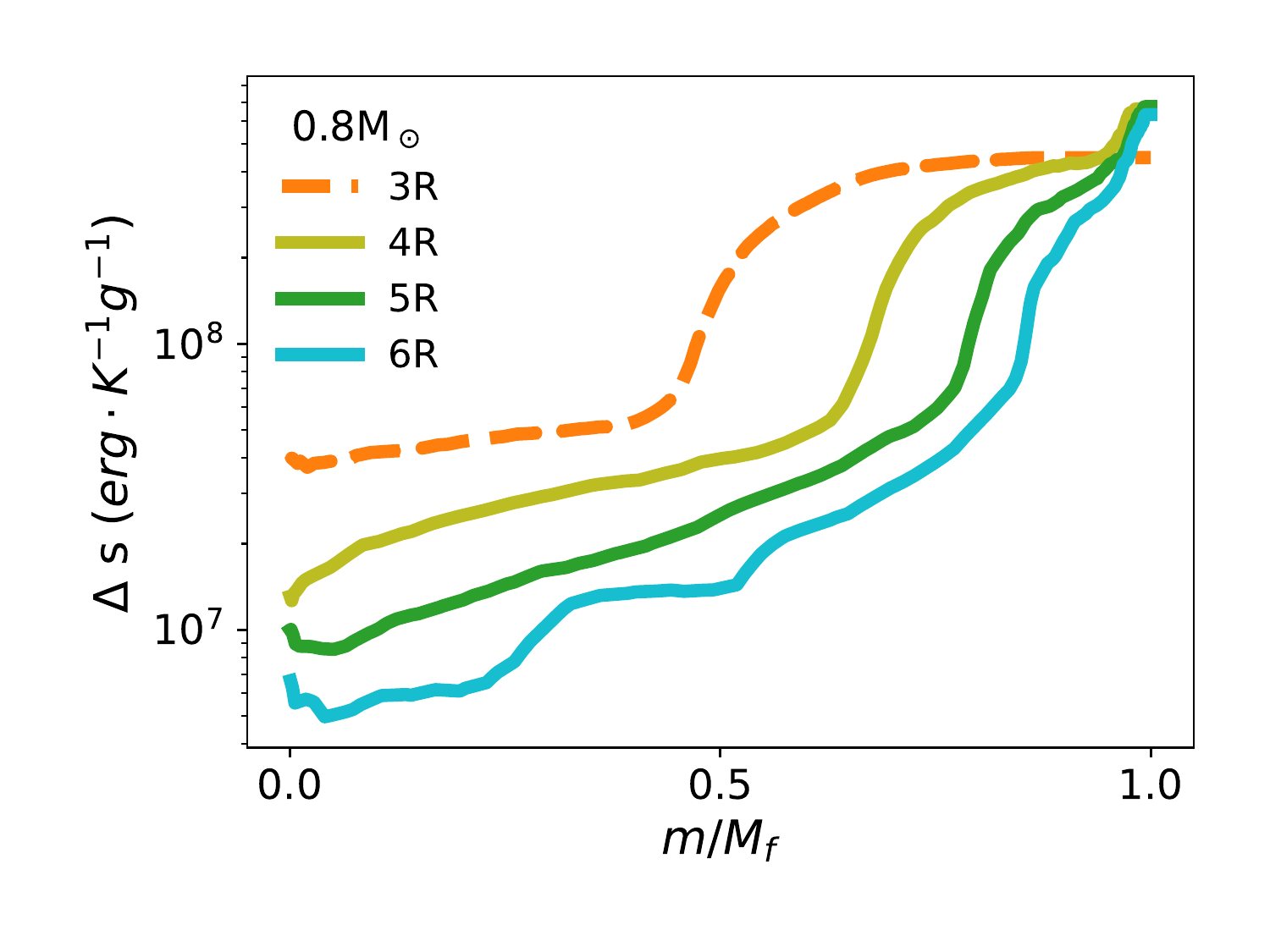}{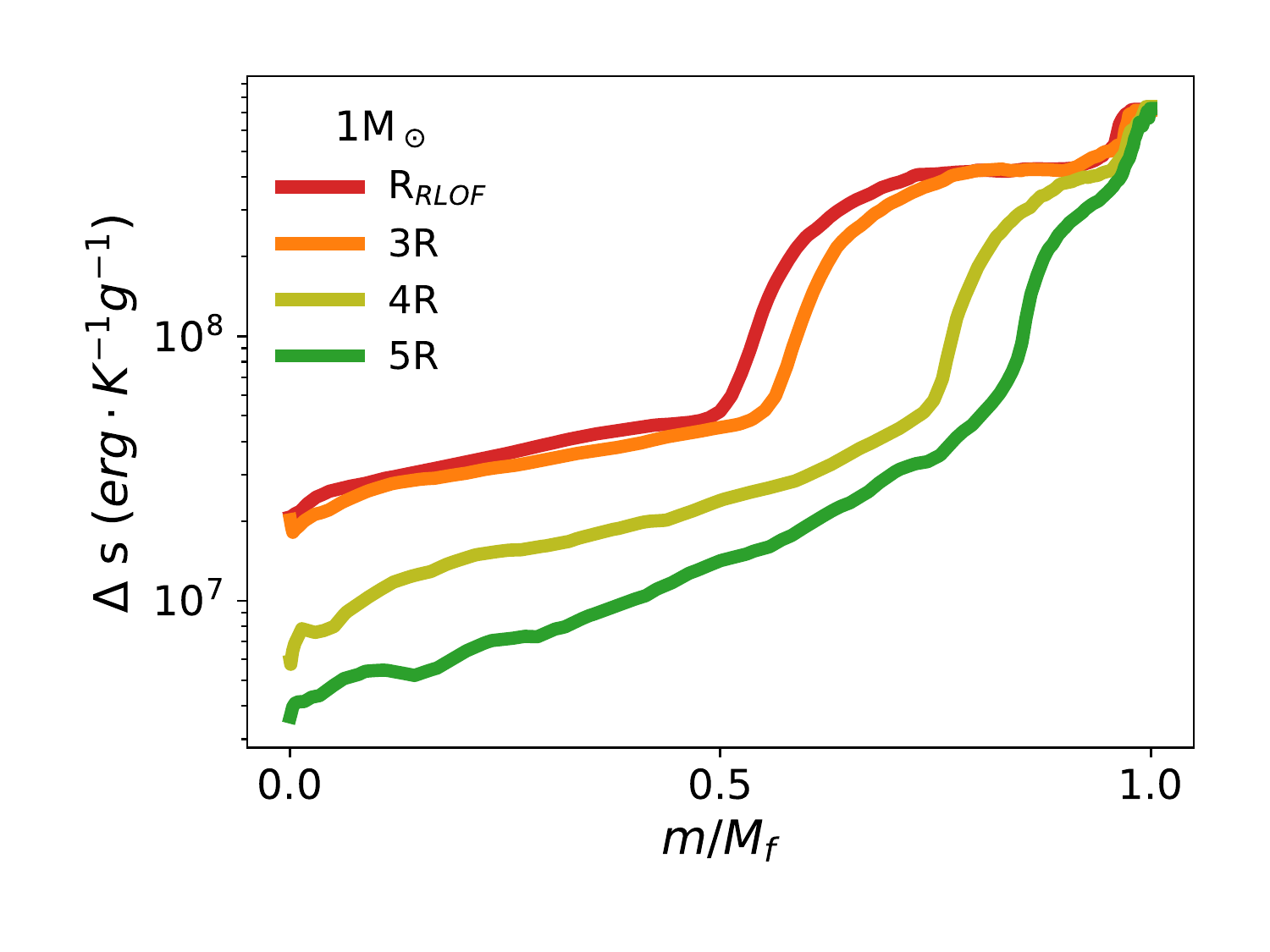}
	\plottwo{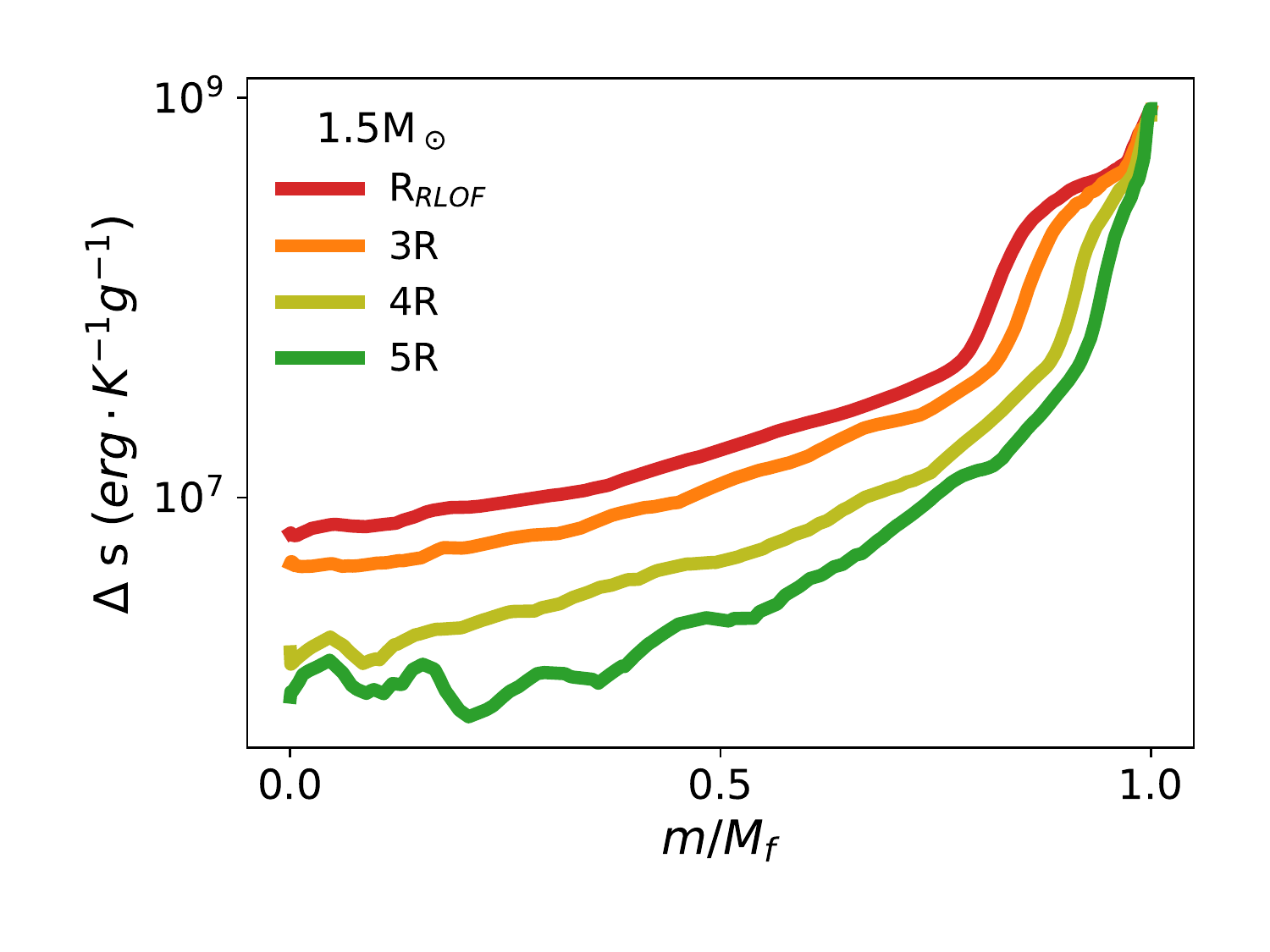}{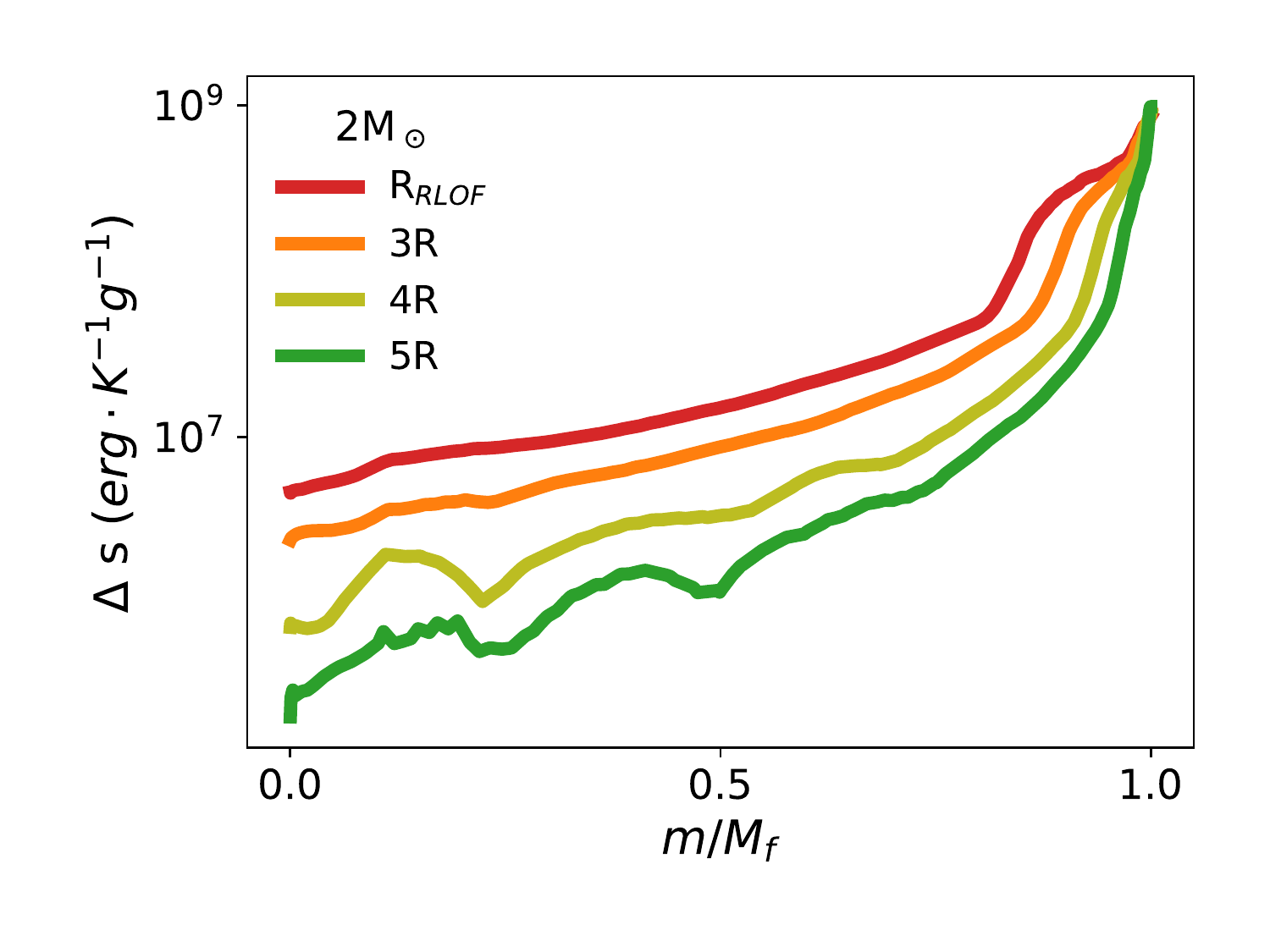}
	\caption{
	Profiles of entropy changes after the SN impact for different models (see Table~\ref{Tab:final}). 
	Since the final unbound mass of model 0.8M3R is still increasing at the end of the simulation (see Figure~\ref{fig_bmass}), we plot it with a dashed line.  
	}
	\label{fig_entro}
\end{figure} 

\begin{figure*}
    \epsscale{0.33}
    \centering
        \plotone{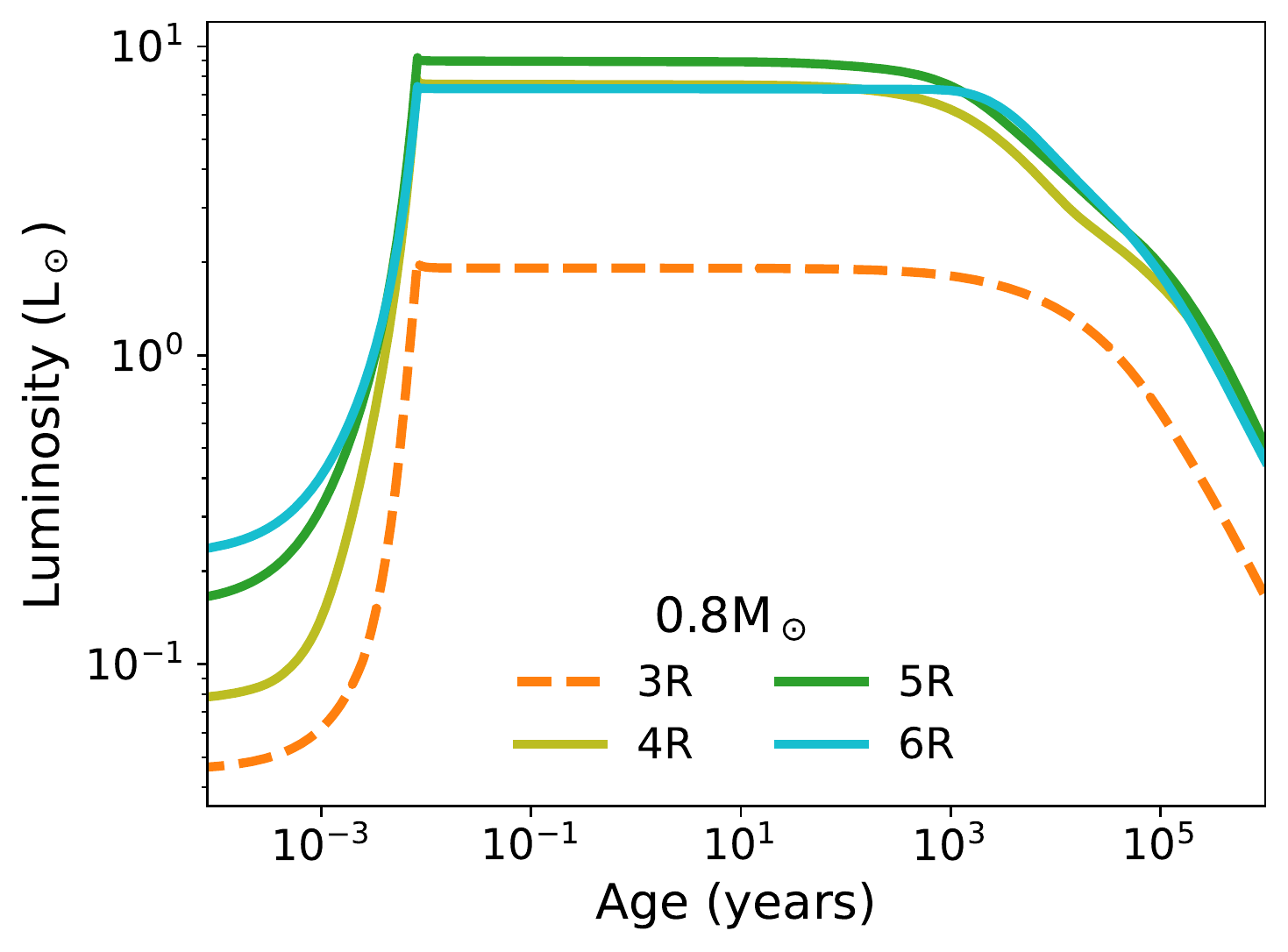}
        \plotone{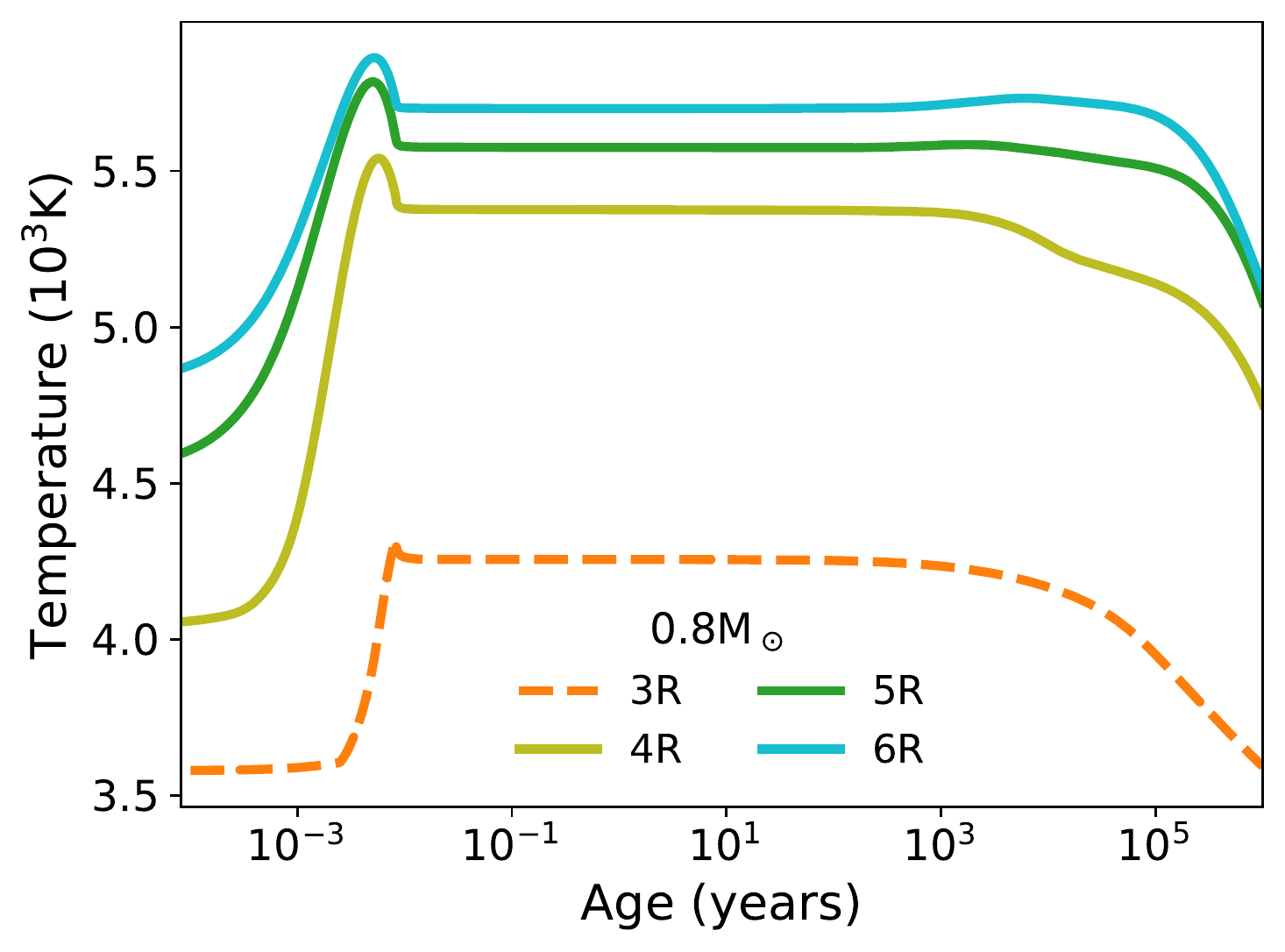}
        \plotone{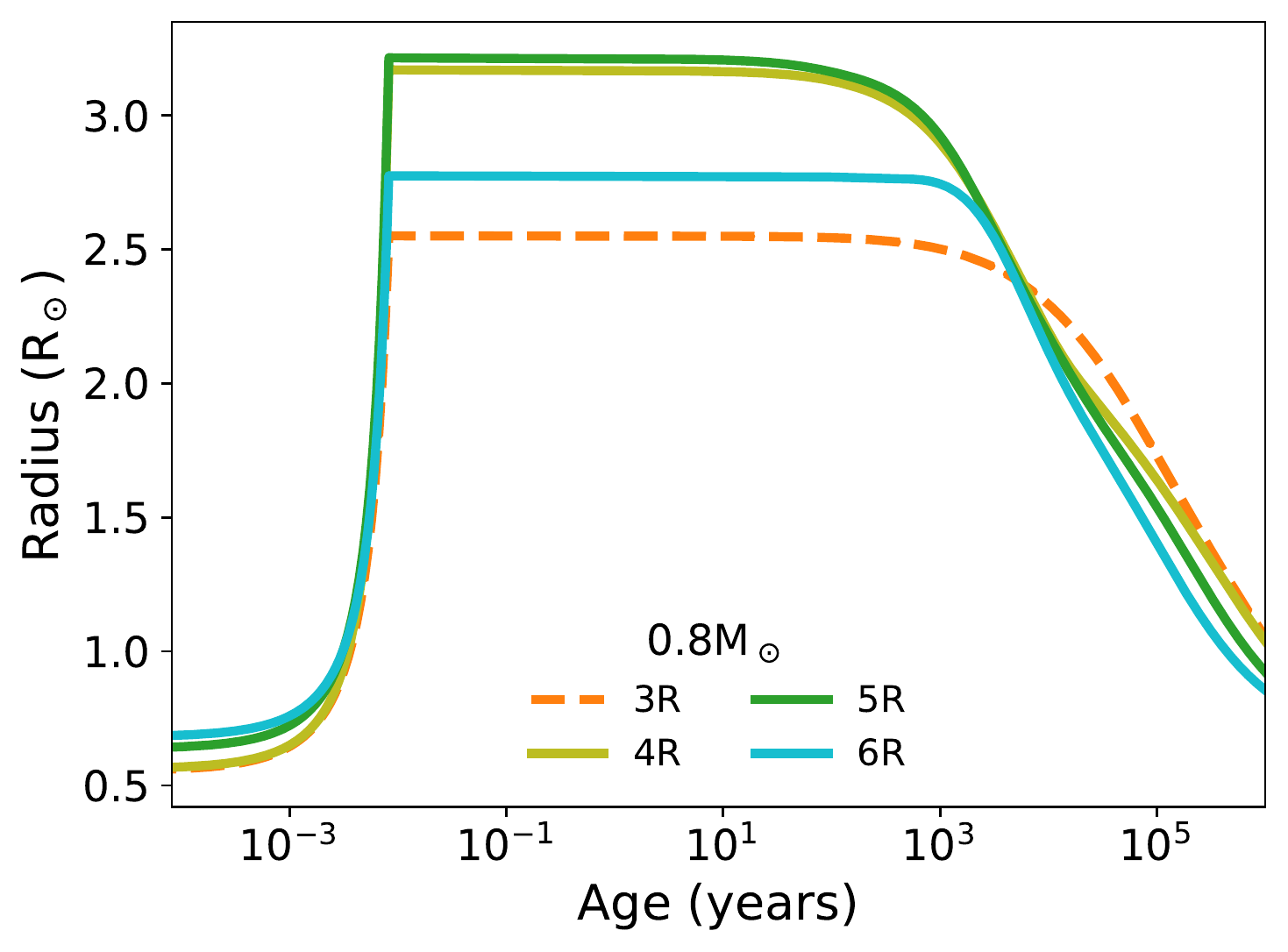}
        
        \plotone{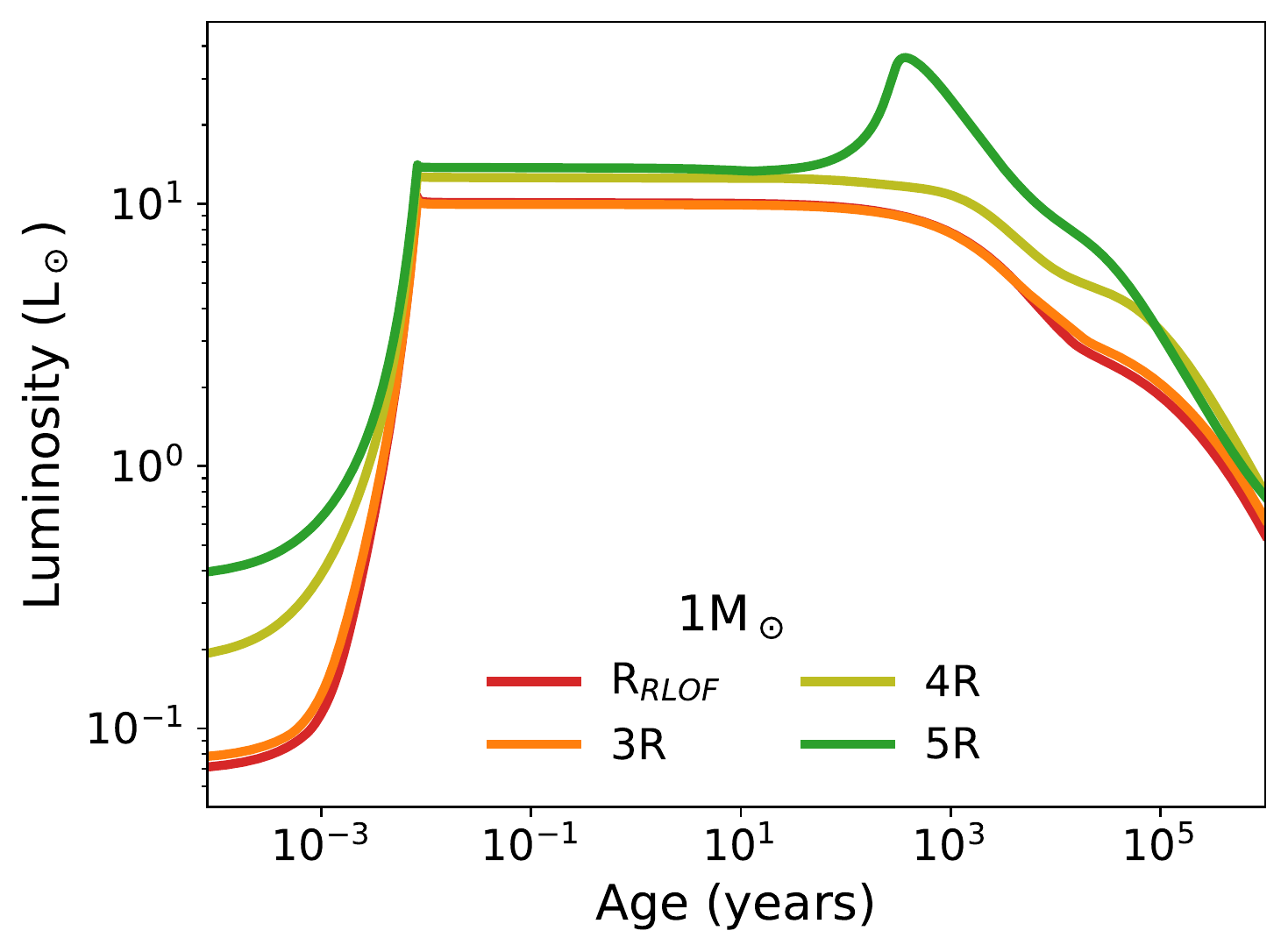}
        \plotone{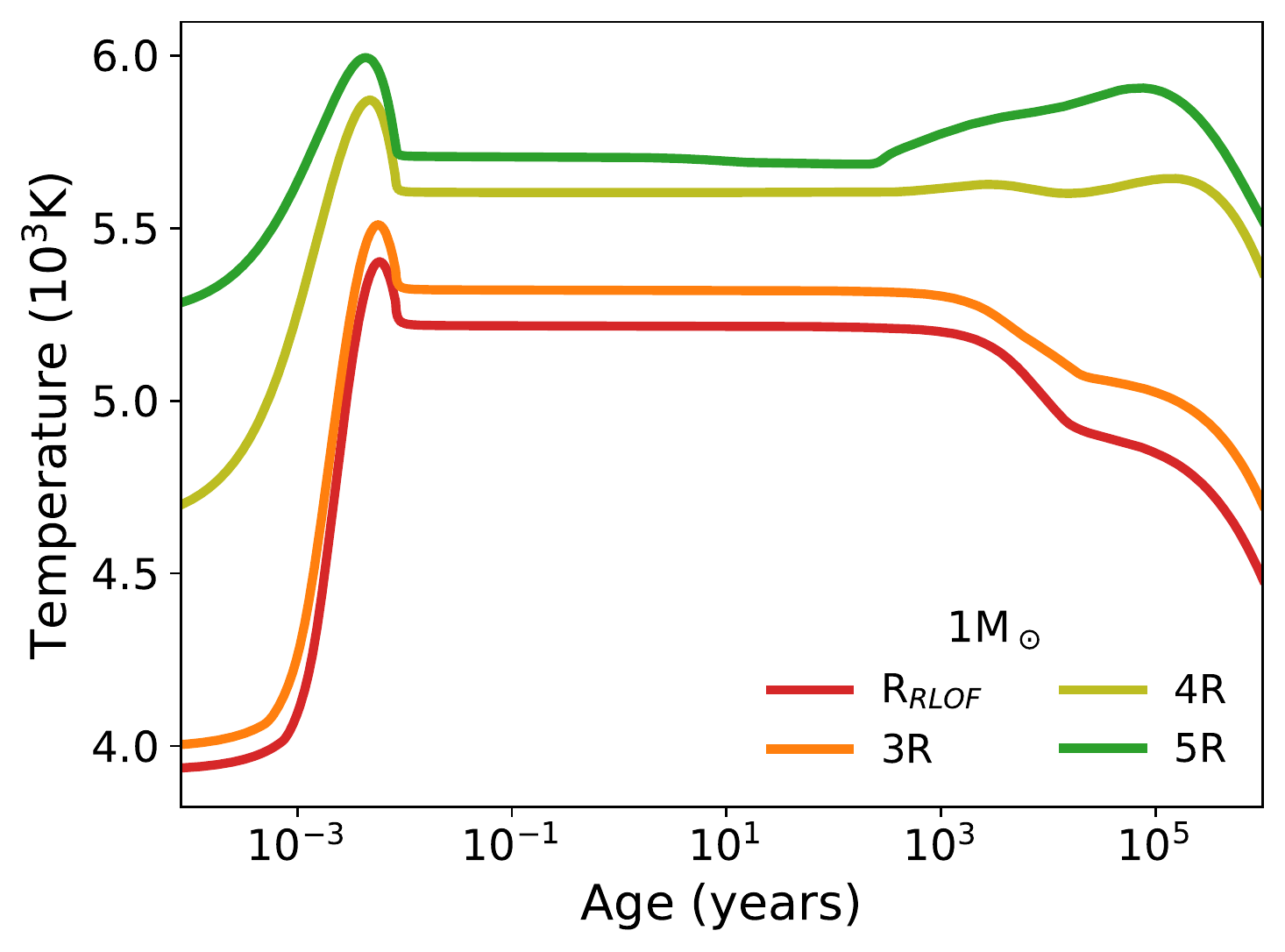}
        \plotone{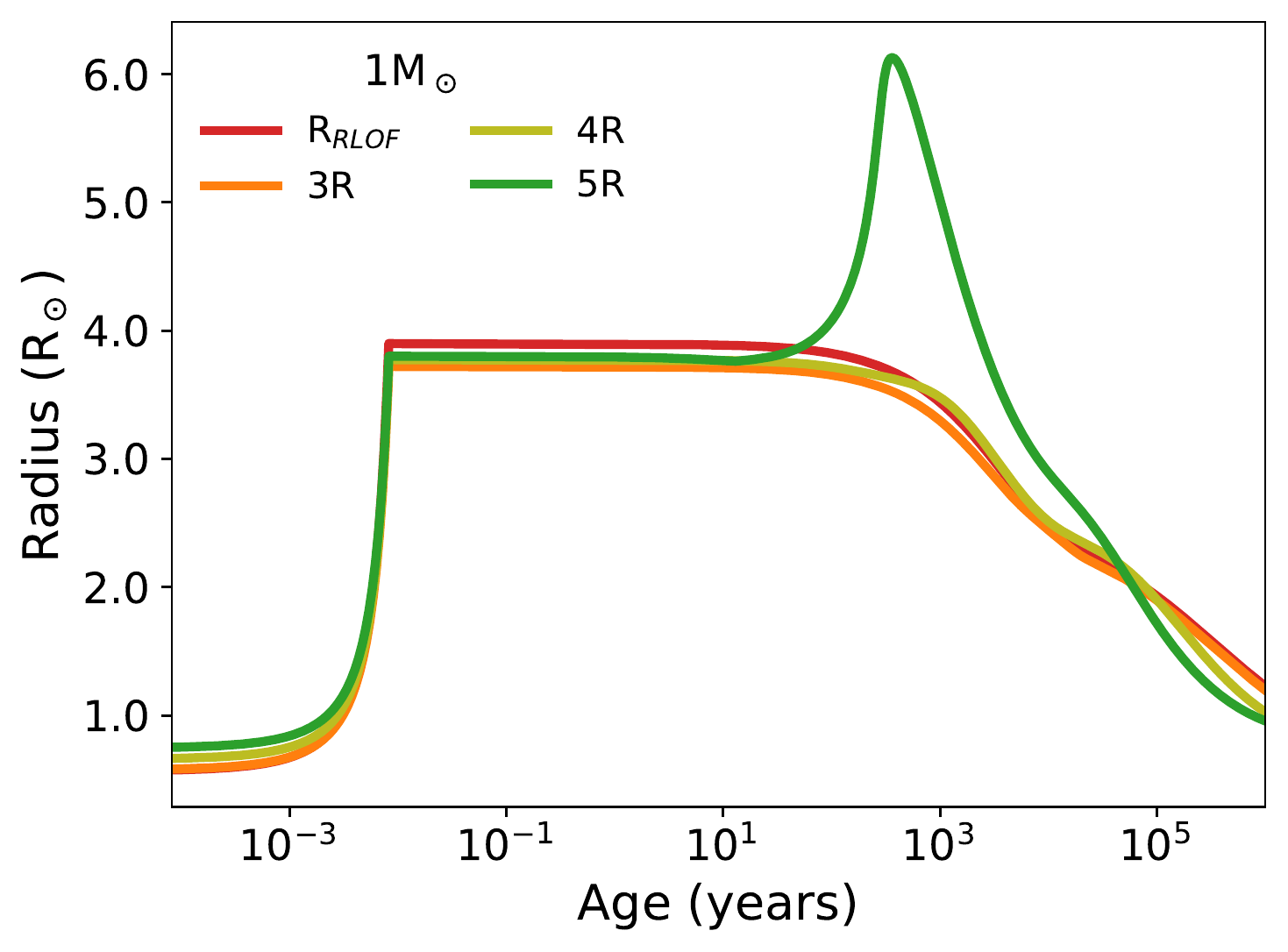}
        
        \plotone{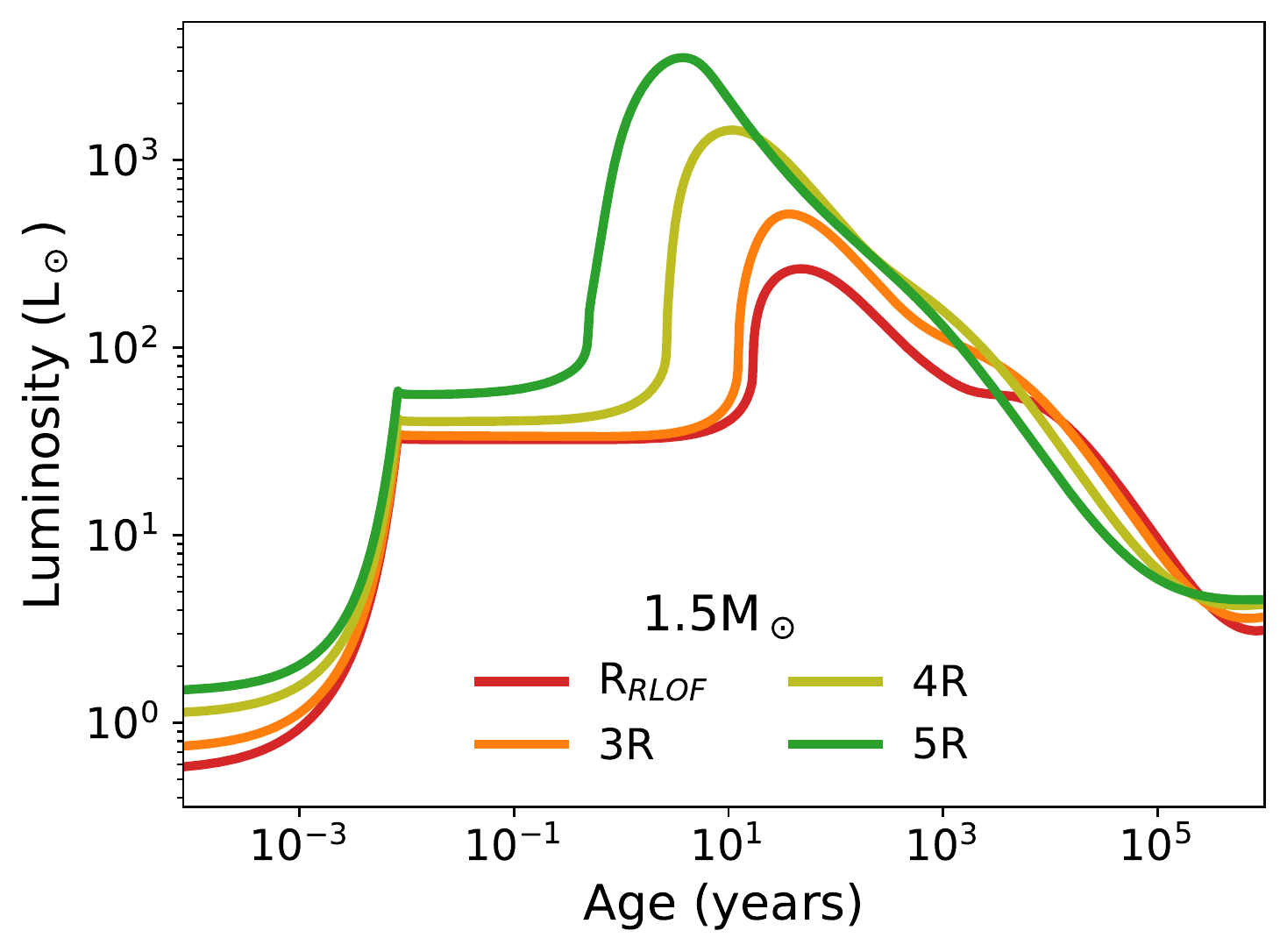}
        \plotone{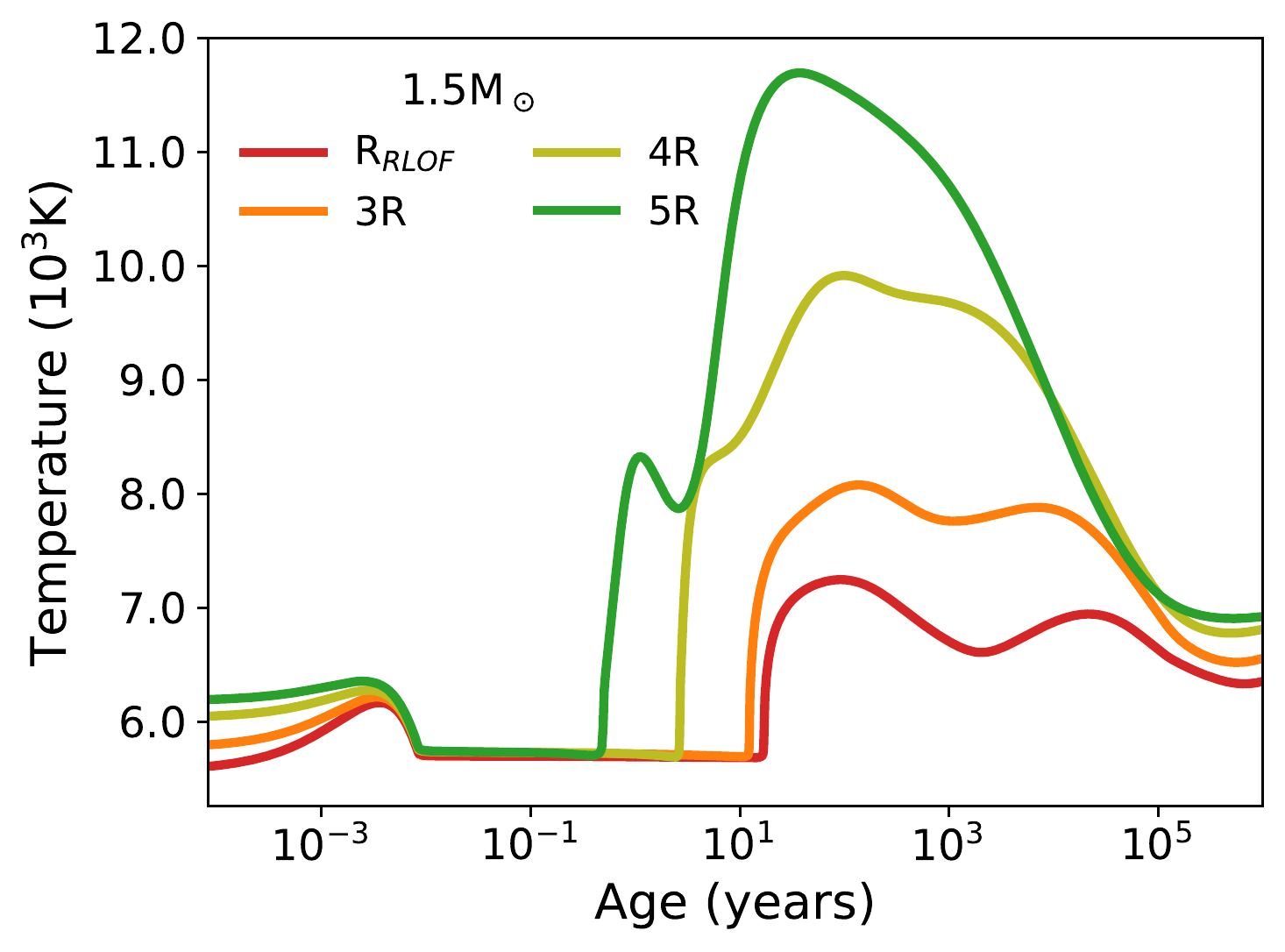}
        \plotone{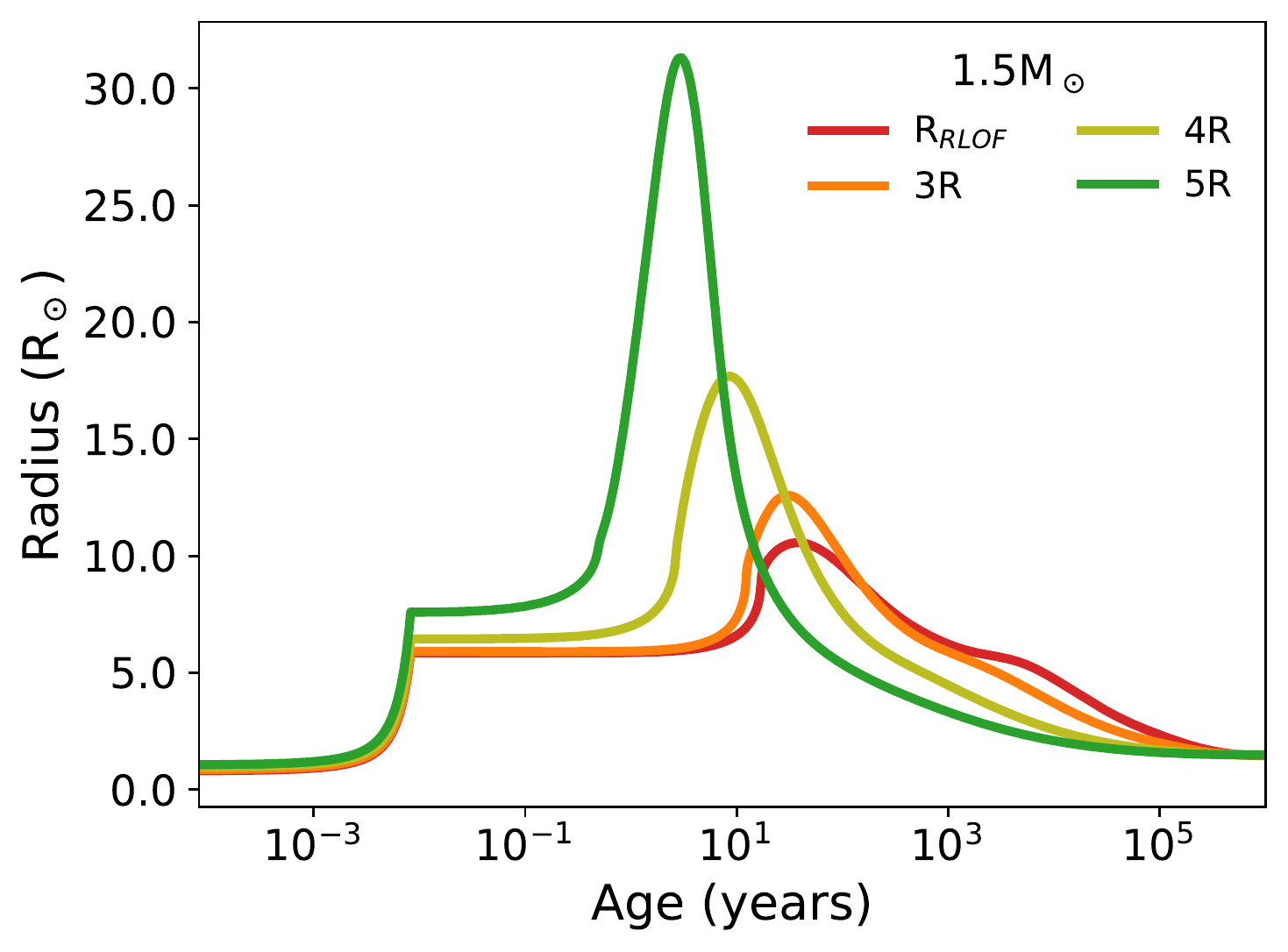}
        
        \plotone{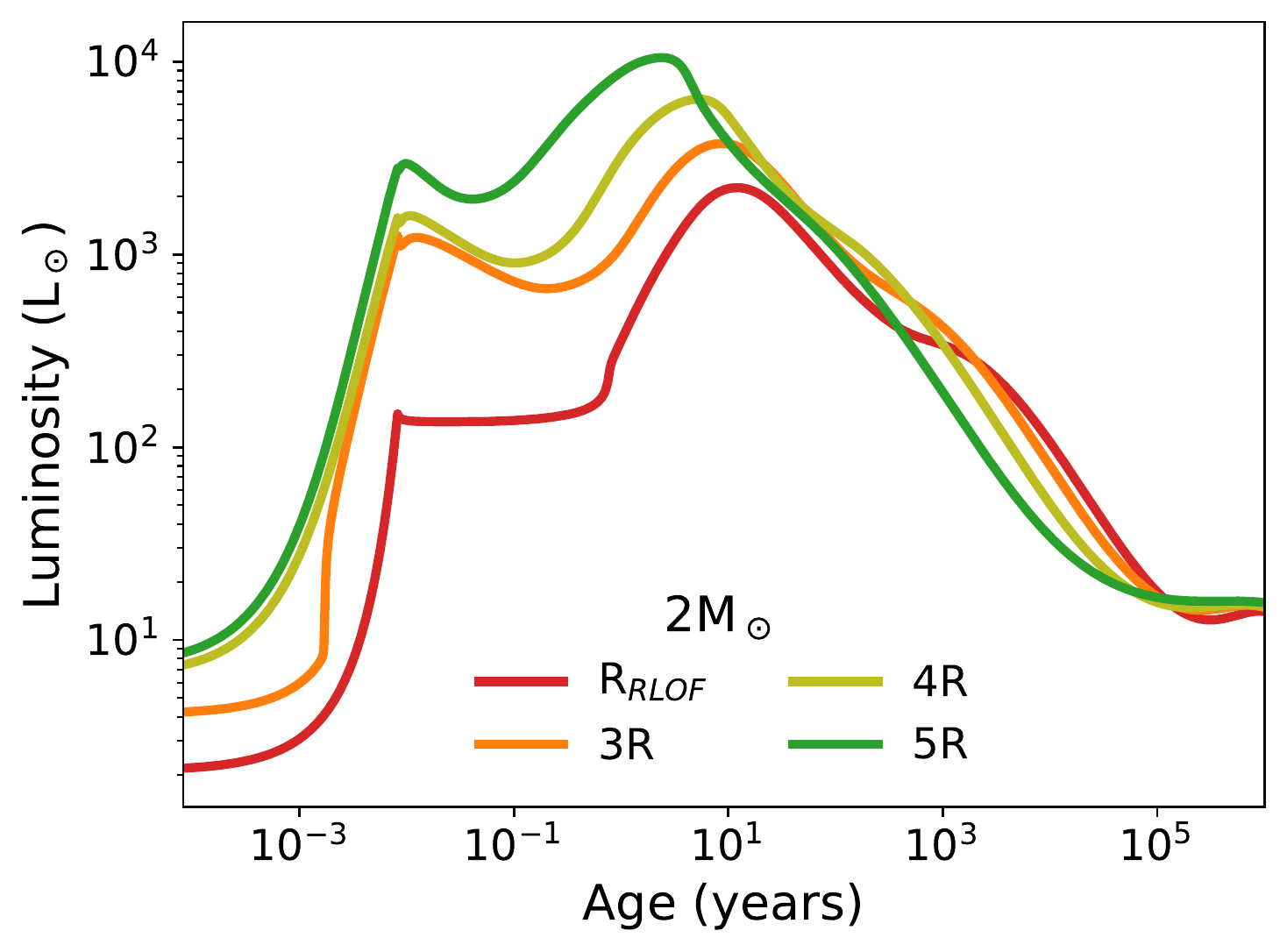}
        \plotone{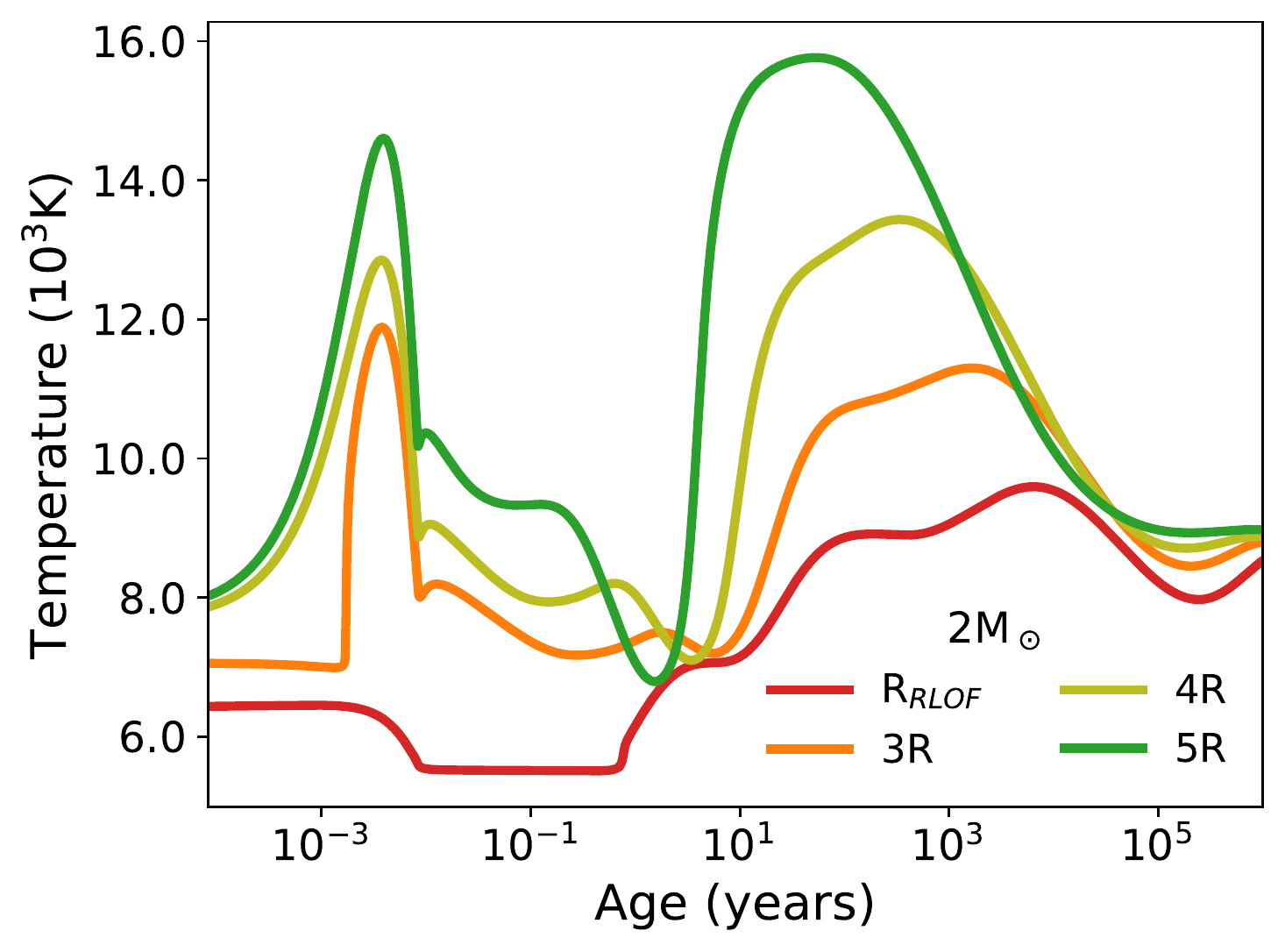}
        \plotone{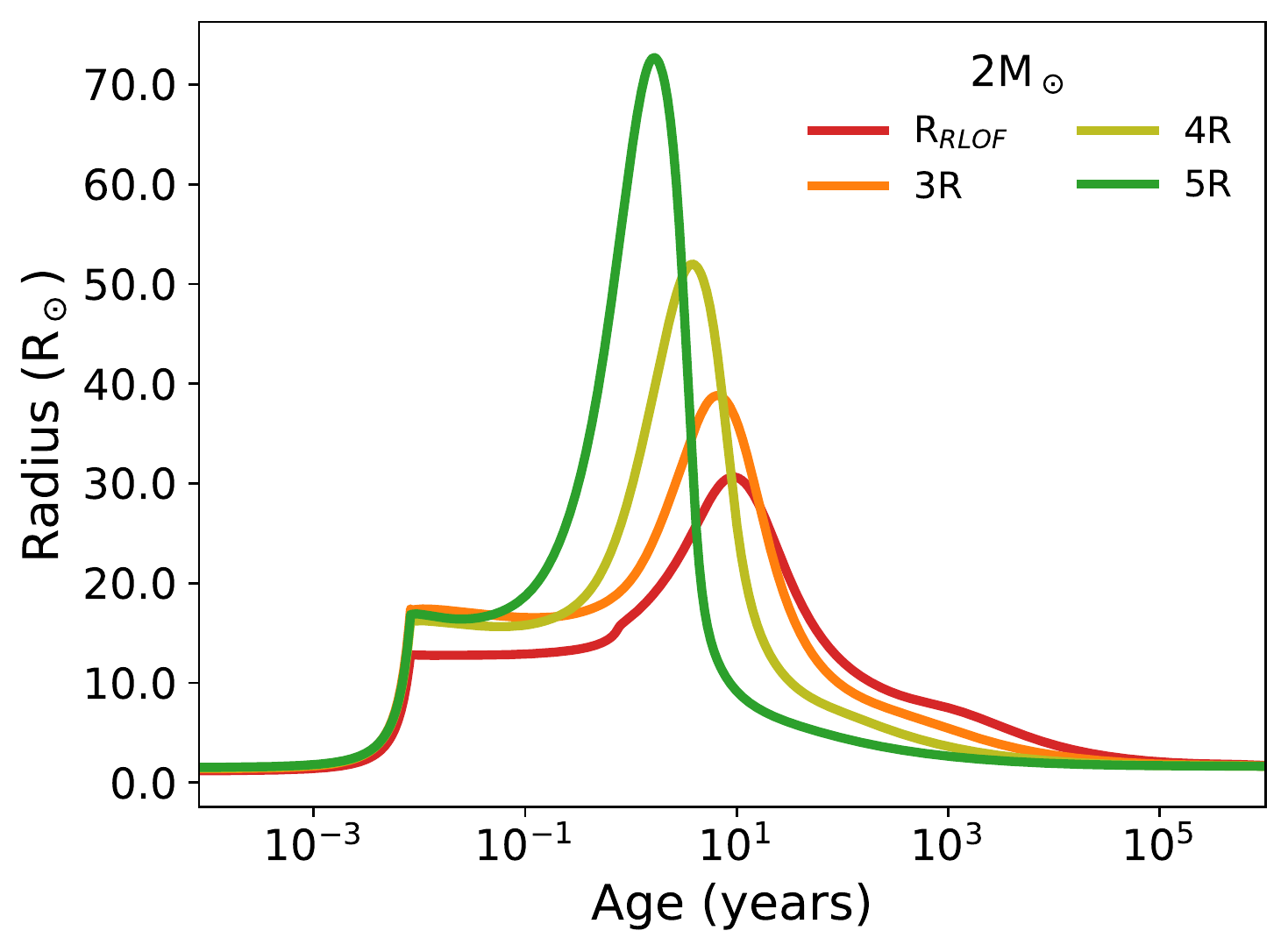}
	
	\caption{
	Evolution of bolometric luminosity, effective temperature, and photosphere radius as functions of time for our models described in Table~\ref{Tab:final}. 
	Different colors represent models with different initial binary separation. Different rows indicates different companion models.  
	}
	\label{fig_evo_sc}
\end{figure*}

%
%
\subsection{Evolution of surviving companions}

As described in Section~\ref{sec:sc}, we measure the mass losses and entropy changes after the supernova impact in {\tt FLASH}
and use them as parametrized heating and mass \kcc{loss} in {\tt MESA}.
Figure~\ref{fig_entro} shows the profiles of the entropy changes for our considered models.  
For higher mass \kcc{loss} models, i.e. models 0.8M3R, 1MRL, and 1M3R, 
the supernova energy could penetrate the companion and significantly heat the companion up to a depth of $0.4 m/M_f$,
where $m/M_f$ is the fraction of the enclosed mass in the mass coordinate.
For more compact companions, such as models 1.5M and 2M, 
the supernova impact contributes heating mainly around the companion surface ($\sim 0.8 m/m_*$). 
The depth of supernova heating determines the local thermal timescale of a surviving companion, 
and therefore affect the time scale of the post-impact surviving companion evolution \citep{2012ApJ...760...21P}.

\begin{figure*}
	\epsscale{0.55}
	\centering
	\plotone{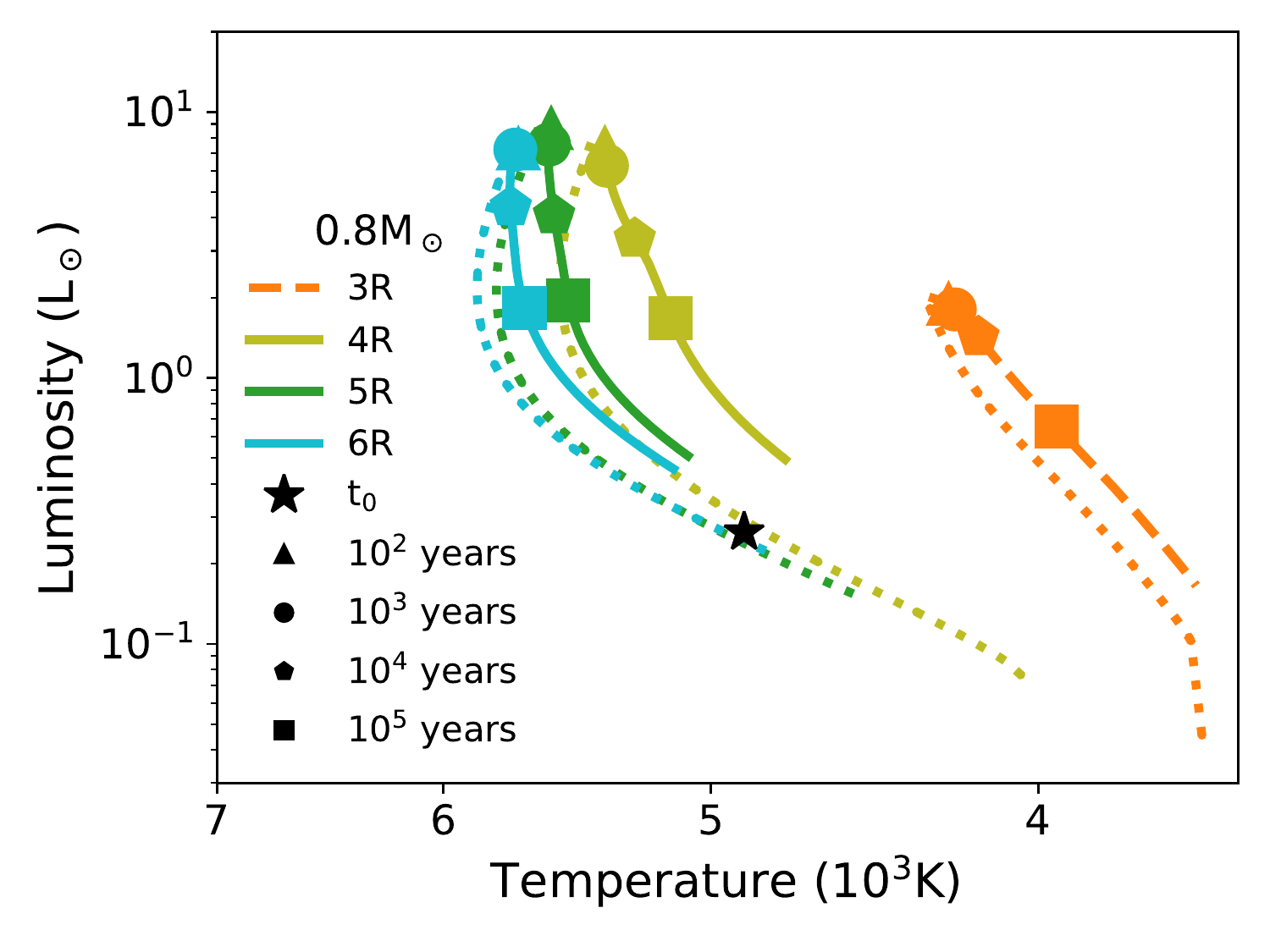}
	\plotone{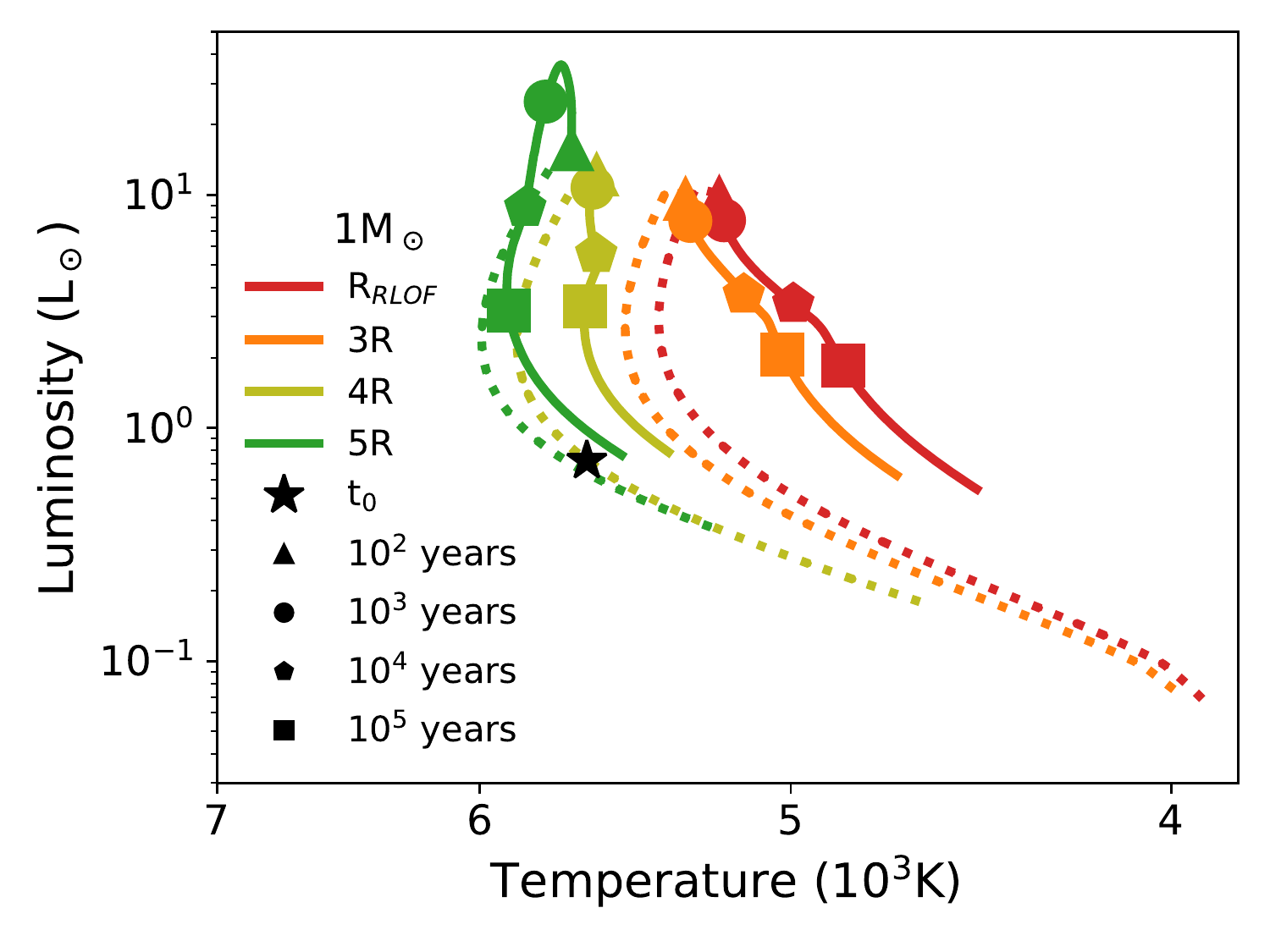}

	\plotone{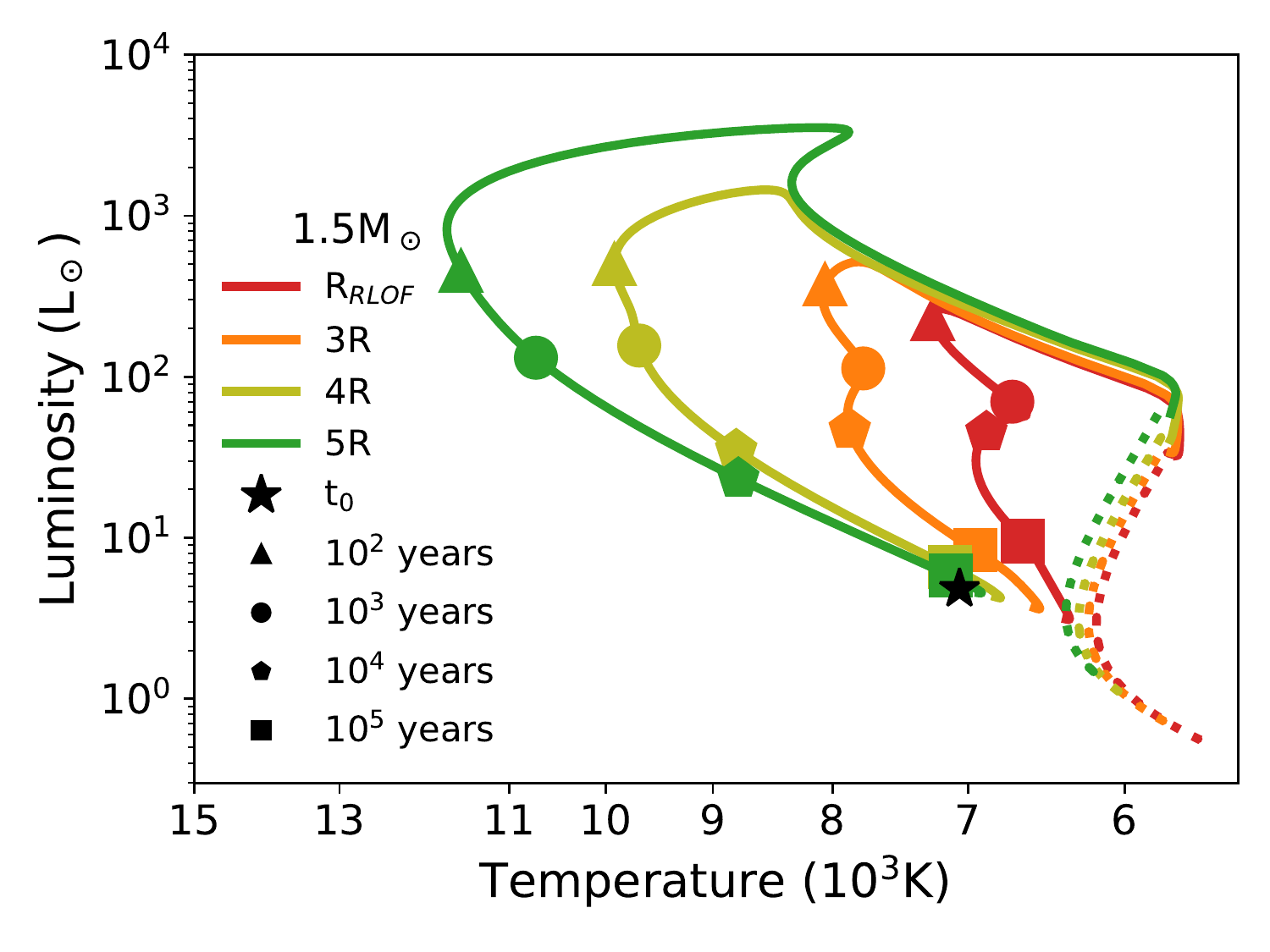}
	\plotone{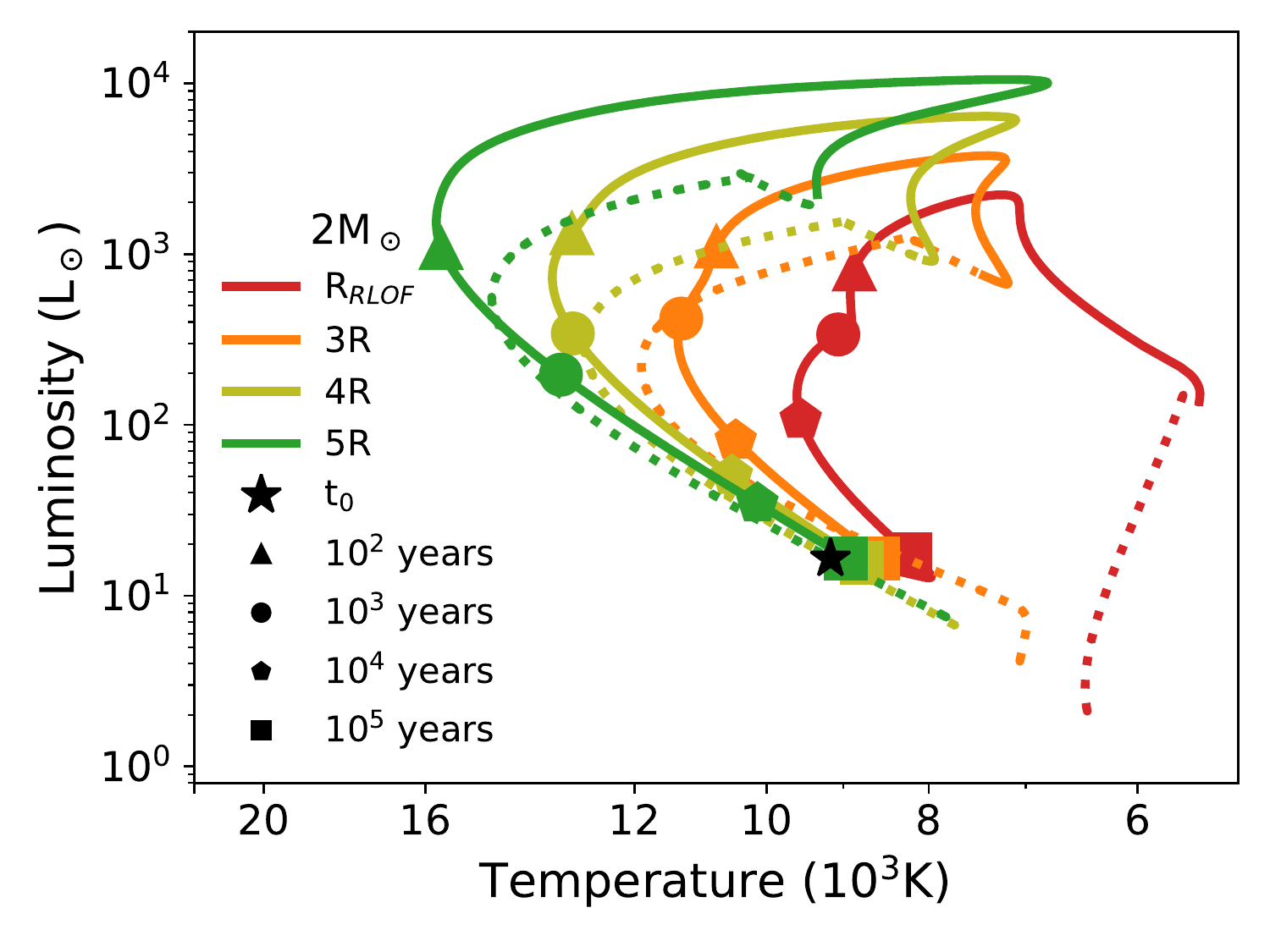}

	\plotone{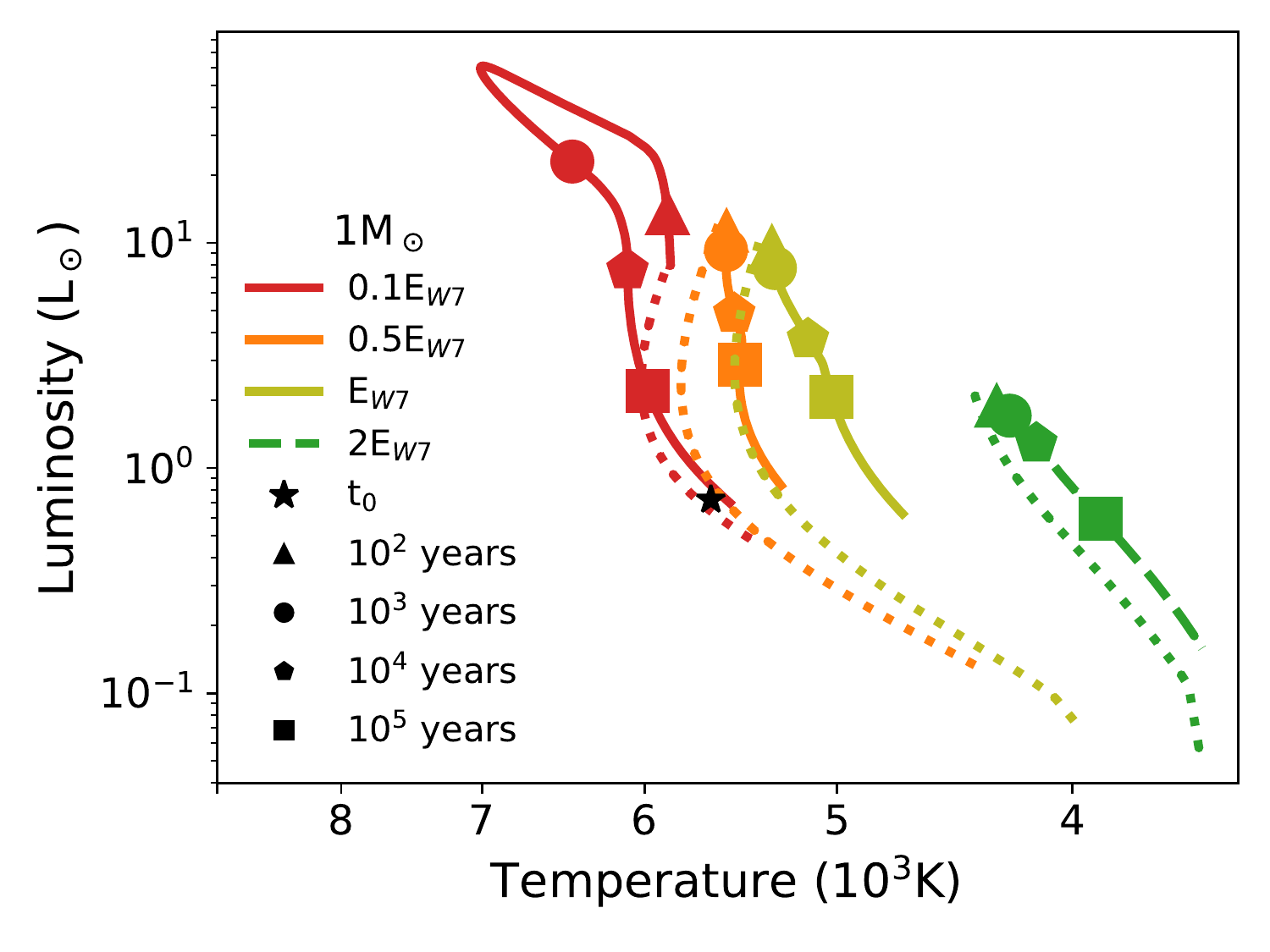}
	\plotone{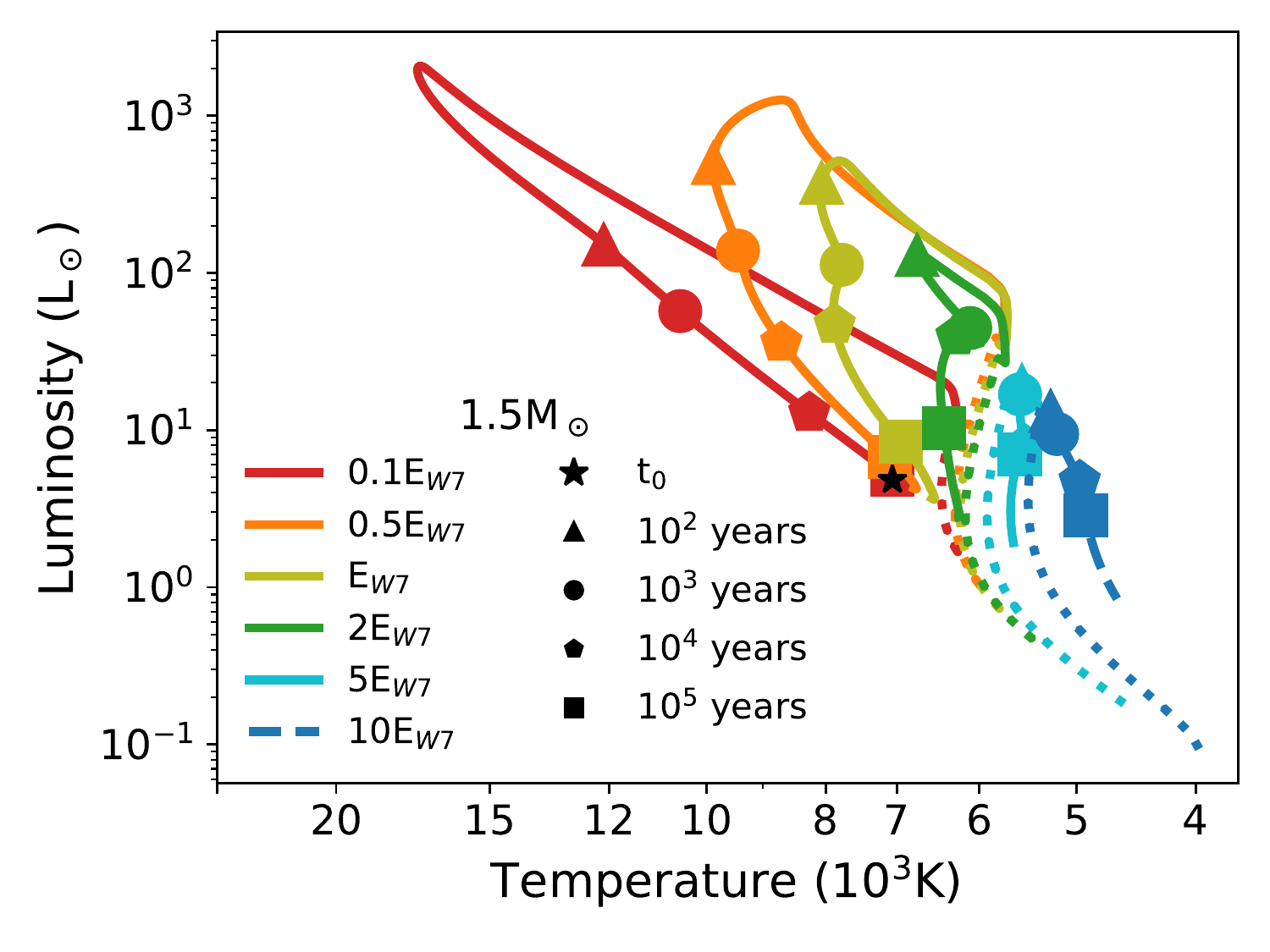}
	
	\caption{
		Evolutionary tracks on the HR diagrams for our models described in Table~\ref{Tab:final}. 
		Different colors represent different initial binary separations \kcc{(the top two rows)} or different explosion energy \kcc{(the bottom row)}.
		We label the times since explosions with different symbols. \kcc{t$_0$ indicates the time of the supernova explosion. 
		Dotted lines present the phase of supernova heating in {\tt MESA}.
		Same reason as in Figure~\ref{fig_entro}, we label 0.8M3R, 1M3RE2, and 1.5M3RE10 model with dashed lines.}
	}
	\label{fig_hr_sc}
\end{figure*}

Figure~\ref{fig_evo_sc} shows the time evolution of the bolometric luminosity, effective temperature, and photosphere radius of our considered models. 
\kcc{The overall features are similar to what have been reported in \cite{2012ApJ...760...21P} with evolved main-sequence companions.
We find that the luminosity ranges and effective temperatures in ours cases with binary separation that just allowed RLOF are matching with the ranges in \cite{2012ApJ...760...21P},
suggesting that the different numerical treatment of SN impact in {\tt MESA} in this study is consistent with the method described in \cite{2012ApJ...760...21P}.} 
The sharp increment at around $10^{-3}$~yrs is due to the sudden heating from the supernova ejecta in {\tt MESA} with a heating timescale $\Delta t=3$~days. 
A second brightening happens at $\sim 10^1 - 10^2$~yrs for models 1M, 1.5M, and 2M, when the deposited energy diffuse away by expanding the photosphere.  
This timescale is described by the local thermal timescale in the stellar envelope \citep{2012ApJ...760...21P}. 
The deeper the energy deposition, the longer the local thermal timescale. 
In addition, models with larger binary separations tend to have shallower energy deposition (see Figure~\ref{fig_entro}). 
Therefore, a more massive companion (less mass stripping) with large binary separation are expected to be over-luminous in young SNRs.
Once the deposited energy is released, the surviving companion starts to contract due to the gravity.

Figure~\ref{fig_hr_sc} shows the corresponding evolutionary tracks of our considered models in HR diagrams. 
We find that the evolutionary tracks can be categorized to two types based on their compactness (similar to the behavior in the final unbound mass): 
Soft companions (models 0.8M and 1.0M) have higher final unbound mass and deeper energy deposition, 
resulting in less luminous surviving companions with $L < 10 L_\odot$.
On the other hand, stiff companions (models 1.5M and 2M) have less final unbound mass and shallower energy deposition, 
leading to a luminous surviving companion with $L > 10^3 L_\odot$ for hundreds of years \kcc{after the explosion time (t$_0$)}. 
The \kcc{bottom} panels in Figure~\ref{fig_hr_sc} presents the evolutionary tracks of surviving companions with different explosion energies.     
Regardless of compactness, stronger the explosion leads to a deeper energy deposition and a larger final unbound mass, 
and therefore produces a dimmer surviving companion.  
For instance, the model 1.5M3R has a peak luminosity $L_{\rm max}\sim 500 L_\odot$ at around 35~years after the SN explosion. 
However, if the SN explosion energy is $\gtrsim 5~E_{W7}$, the peak luminosity could drop to $L_{\rm max} < 23 L_\odot$ in Figure~\ref{fig_hr_sc}. 
The current non-detection of surviving companions in nearby SNRs might be explained by a soft surviving companions, 
shorter binary separations due to the delayed explosion, or super-Chandrasekhar mass explosion with a higher explosion energy.
In latter case, the SN impact will be strong than a typical binary separation by RLOF, which might show stronger UV excess in the early lightcurve \citep{2010ApJ...708.1025K}.

\section{Summary and Conclusions \label{sec4}}

We have performed 2D hydrodynamics simulations of supernova impact and the subsequence 1D stellar evolution of surviving companions 
with a wide range of parameters of main-sequence like companions. 
\kcc{It should be noted that the simulations presented are 2D, which ignored both spin and orbital motions, and with non-evolved binary companions. 
\cite{2012ApJ...750..151P} suggest that the spin and orbital motions could make some difference during the SN impact and therefore might affect the later on surviving companion evolutions (but less important as suggested by \citealt{2021A&A...654A.103L}).}
\kcc{Even though, we find three universal relations (Equations~\ref{eq_bmass}, \ref{eq_vkick}, and \ref{eqn: Dheat}) 
to describe the final unbound mass, kick velocity, and SN heating depth as functions of binary separation, 
regardless the companion mass or evolution stage in MS.}
Stiff surviving companions with larger progenitor masses (models 1.5M and 2M) have less mass stripping during the SN impact and received shallower supernova heating in the envelope.
The deposited energy in the envelope will be diffused away within hundreds of years years, letting the photosphere expands.
This type of surviving companions will be over-luminous ($L_{\rm max} > 10^2 - 10^4 L_\odot$) and have higher chance to be detected in nearby SNRs. 
In contrast, soft surviving companions with lower progenitor masses (models 0.8M and 1M) result in larger final unbound mass and deeper supernova heating.
These low mass surviving companions could remain $L_{\rm max} < 10L_\odot$, which could explain the non-detection of surviving companions in nearby SNRs. 
In addition, super-Chandrasekhar mass explosion with a higher explosion energy or a shorter binary separation during the spin-up/down phase could produce low luminosity surviving companions.  
In these cases, the surviving companions will have a higher kick velocity, 
and the early UV or X-ray light from the collision between the SN ejecta and companion will be stronger than the previous estimate in \cite{2010ApJ...708.1025K} for MS-like companions.

\acknowledgments
\kcc{We thank the anonymous referee for his/her valuable comments and suggestions that helped to improve this paper. }
SJR and KCP acknowledge valuable discussions with You-Hua Chu and Hsiang-Yi Karen Yang on the searches for surviving companions. 
\kcc{SJR also thanks Dr.~Stephen Justham for his useful comments and suggestions during her virtual visit in IMPRS.}
This work is supported by the Ministry of Science and Technology of Taiwan through grants MOST 107-2112-M-007-032-MY3 and MOST 110-2112-M-007 -019,
by the Center for Informatics and Computation in Astronomy (CICA) at National Tsing Hua University through a grant from the Ministry of Education of Taiwan.
{\tt FLASH} was in part developed by the DOE NNSA-ASC OASCR {\tt Flash} Center at the University of Chicago.
The simulations and data analysis have been carried out at the {\tt Taiwania} supercomputer in the National Center for High-Performance Computing (NCHC) in Taiwan, 
and on the CICA cluster at National Tsing Hua University.
Analysis and visualization of simulation data were completed using the analysis toolkit {\tt yt}.

\software{MESA 10398 \citep{MESA11, MESA13, MESA15, MESA18, MESA19}, FLASH 4.5 \citep{FLASH4}, {\tt yt} \citep{2011ApJS..192....9T}, and {\tt python-helmholtz (https://github.com/jschwab/python-helmholtz)} \citep{helm1, helm2}}.

%
%
%
%



\end{document}